\documentclass[10pt]{IEEEtran}
\usepackage{amsmath}
\usepackage[T1]{fontenc}
\usepackage{algorithm}
\usepackage{alphalph}
\usepackage{algorithmic}
\usepackage{listings}

\usepackage{nomencl}
\makenomenclature

\usepackage{amsmath}
\usepackage{graphicx}
\usepackage{color}
\usepackage[cmintegrals]{newtxmath}
\usepackage[caption=false,font=footnotesize]{subfig}
\ifCLASSOPTIONcompsoc
\usepackage[caption=false,font=normalsize,labelfont=sf,textfont=sf]{subfig}
\else
\usepackage[caption=false,font=footnotesize]{subfig}
\fi
\usepackage[colorlinks,citecolor=red,urlcolor=blue,bookmarks=false,hypertexnames=true]{hyperref} 
\usepackage{caption}
\usepackage{lipsum}
\usepackage{nameref}
\usepackage{varioref}
\usepackage{hyperref}
\usepackage{setspace}

\newtheorem{lemma}{Lemma}
\newtheorem{proposition}{Proposition}

\begin{document}

\title{\textcolor{black}{A Game Theoretical Analysis on the \textit{Gray-Wyner} System and Generalised Common Information}}
\author{Makan~Zamanipour,~\IEEEmembership{Member,~IEEE}
\thanks{Copyright (c) 2015 IEEE. Personal use of this material is permitted. However, permission to use this material for any other purposes must be obtained from the IEEE by sending a request to pubs-permissions@ieee.org. Makan Zamanipour is with Lahijan University, Shaghayegh Street, Po. Box 1616, Lahijan, 44131, Iran, makan.zamanipour.2015@ieee.org. }
}
\markboth{Journal of \LaTeX\ Class Files,~Vol.~xx, No.~x, Xxxx~ 2021}%
{Shell \MakeLowercase{\textit{et al.}}: Bare Demo of IEEEtran.cls for IEEE Transactions on Magnetics Journals}
\maketitle
\IEEEdisplaynontitleabstractindextext

\begin{abstract}
We analyse \textcolor{black}{the common information problem for the generalised \textit{Gray-Wyner} problem}. We aim to explore the problem-and solution in relation to the non-orthogonality among the \textcolor{black}{source decoder}s’ components. We consider a simple networked control system consisting of 2 groups of users: (\textit{i}) one sender or Observer named \textit{Alice}; and (\textit{ii}) a group of multiple receivers or Controllers, named \textit{Bob}s. In order to tackle the possible risk \textcolor{black}{arisen} from the common information among Bobs, Alice provides a redundancy creating some virtual messages which are in the null of each specific Bob, but not for others. The aforementioned possible risk is inevitable since, non-impossibly speaking, some/all of them may instantaneously act as potential Eavesdropper(s) with the abuse of the aforementioned common information. This novel discipline, which has not been investigated yet to the best of our knowledge, is theoretically interpreted from a \textit{mirror-game}-theoretical point-of-view. Novel mathematical problems are derived specifically including some proofs for the information-theoretic relaxations and non-stationarity as well as the existence of the \textit{Nash} equiblirium. Finally speaking, simulations approve our scheme. 
\end{abstract}

\begin{IEEEkeywords}
Anytime capacity, common information, \textit{Concentration-of-Measure} inequalities, controllability, information asymmetry, inner-loop stability, \textit{Lohe} model, max $\mathbb{K}-$kut game, mean field game, \textit{Nash} equilibrium, non-stationarity, Oscillators, rate-distortion, Stackelberg game, uncertainty, virtual twins, \textit{worst case} method.   
\end{IEEEkeywords}

\maketitle

\IEEEdisplaynontitleabstractindextext
\IEEEpeerreviewmaketitle

\section{Introduction} 

Two terms control and information are \textcolor{black}{interchangeable} \cite{1, 2, 3, 4, 5}. The information-theoretic bounds achieved through simultaneous resource use have a lot of priorities compared with the recent random access strategies. This is because of the fact that in reality, controlled networked systems are heterogeneous as they are disrupted by uncertainties. 

\textit{''Does there exist an algorithm that stops in an acceptably finite time zone and outputs an approximation of it, an \textit{inference} would be of an acceptably interpretive nature''}, Shannon source coding theory says. One of the fundamental aims of source coding in network information theory is to quantify how much sources are of an informative $-$ interpretive $-$ nature to each other for the multi-user communication problems. Recent research has proven that this informativity and interpretability depends on the actual application and/or setup, and the bad news is that there is no tight universal notion of information. The most common notion is Shannon’s mutual information, which is the reduction in the entropy of a random variable due to the knowledge of a
correlated random variable.

In some cases in real-time scenarios, it is required the information sources to be simultaneously accessible to multiple legitimate users $-$ e.g. in the \textit{Gray-Wyner} scheme \cite{1}. This issue, however, unhesitatingly creates some potential risks \textcolor{black}{due to an open nature of the system in the context of eavesdropping}.

\subsection{Literature review}
\subsubsection{\textcolor{black}{Information-theory \& control-theory}} From an information-theoretic control-theoretic point of
view, in joint control and network design, the application is
control \cite{2, 3, 4, 5}. In this fashion, some brilliant activities have
been performed. 

\subsubsection{\textcolor{black}{Information theory \& game-theory}} From an information theoretic game-theoretical point of
view, although some \textcolor{black}{few} work has been done e.g. \cite{3, 4}, this dominant area of research is still
open. In \cite{3}, a trade-off was realised in the context of ”how much
fast and how much secure” according to two essential limits in mean-field-games. In \cite{4}, the information
pattern of a mean-field-theoretical scheme was evaluated.

\begin{figure*}[t] 
\centering
\subfloat{\includegraphics[trim={{14mm} {93 mm} {227mm} {17mm}},clip,scale=0.5]{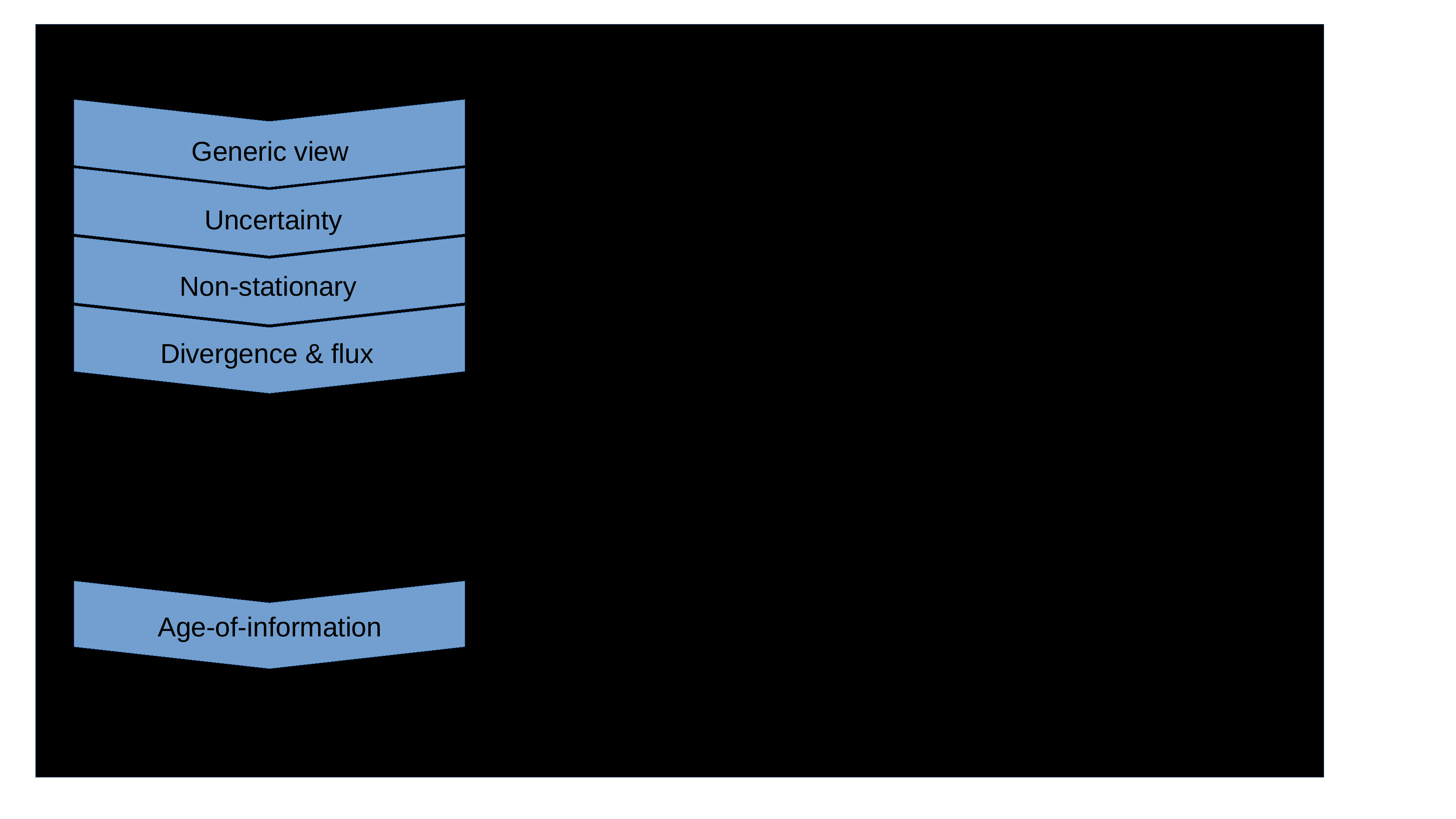}}\\
\caption{\textcolor{black}{Flow of the problem-and-solution throughout this paper}.} \label{F0}
\label{fig:EcUND} 
\end{figure*} 

\subsubsection{\textcolor{black}{\textit{Gray-Wyner}'s common information}} In the fashion of the common information problem for the \textit{Gray-Wyner} framework, some few work has been fulfilled. In \cite{6}, a two-step strategy was proposed for symmetric sources which would guarantee the lower bound for large cache capacities. \textcolor{black}{Meanwhile, it was
proven that, conditioned on the third source, it would be
satisfied during the half of the joint entropy of the two sources.} In \cite{7}, it was shown that where hardly are exact cache hits satisfied in the real-time media, network cached information would be inevitably pivotal for network compression due to the entanglement. In \cite{8}, more novel insights were added to \cite{6} from a theoretical point of view achieving some new bounds. In \cite{9}, non-causality and lossy issues were explored for the generalised Gray-Wyner set ups. In \cite{10, 11}, some new conditions were evaluated over the joint statistics of the pair of sources. \textcolor{black}{The equality of Wyner and exact common information was consequently proven for the generalised erasure and binary Z-sources.}

\subsubsection{\textcolor{black}{Generalised common information}} \textcolor{black}{In relation to the generalised common information in the generalised \textit{Gray-Wyner} schemes, some works have also been perfectly presented in the literature such as \cite{12, 13, 14, 15, 16, 17, 18, 19, 20}. However, the distortion bounds derived in the aforementioned work for source broadcast issues with the generalised common information were of a purely information-theoretic nature. }\textcolor{black}{More specifically, in \cite{12}, the issue of multi-user privacy for the Gray-Wyner’s generalised common information was fully explored. In \cite{13} and for the generalised common information, the principle of measuring commonness through the conditional maximal correlation was theoretically investigated. In \cite{14},  a novel lossy source coding analysis was fully performed for the aforementioned systems. In \cite{15}, the total correlation of Gaussian vector sources for the aforementioned systems was analysed as well. In \cite{16, 17}, the issue of rate-distortion region of Gray-Wyner frameworks with helpers was theoretically evaluated. In \cite{18}, for a tuple of correlated multivariate Gaussian RVs, a novel lossy network compression method was newly proposed. In \cite{19, 20}, some information-theoretic metrics were theoretically explored for the problem of the Gray-Wyner’s generalised common information. }


\subsection{Motivations and contributions}
In this paper, we are interested in responding to the following questions: \textit{What if a network generally is under a potential risk? How can we explore it from a game-theoretical point of view, so as to control of it? Is it possible for us to enjoy the common information rather than being annoyed of it? Is it possible to find a relaxed solution to the main problem as much as possible? Can we find a Nash equilibrium for our game theoretical paradigm? What if our control-input is non-stationary? Can we solve the main problem from a game theoretical point of view? What if a Major-Player influences the overall throughput in our game scenario? What if this Major-Player is non-stationary as well? What other solutions do we get access to?} With regard to the incomplete version of the literature $-$ although some tight bounds have been theoretically derived from an information-theoretic standpoint $-$ , the questions expressed here strongly motivate us to find an interesting solution, according to which our contributions are described in the following.

\begin{itemize}

\begin{figure*}[t] 
\centering
\subfloat{\includegraphics[trim={{74mm} {39 mm} {83mm} {44mm}},clip,scale=0.38]{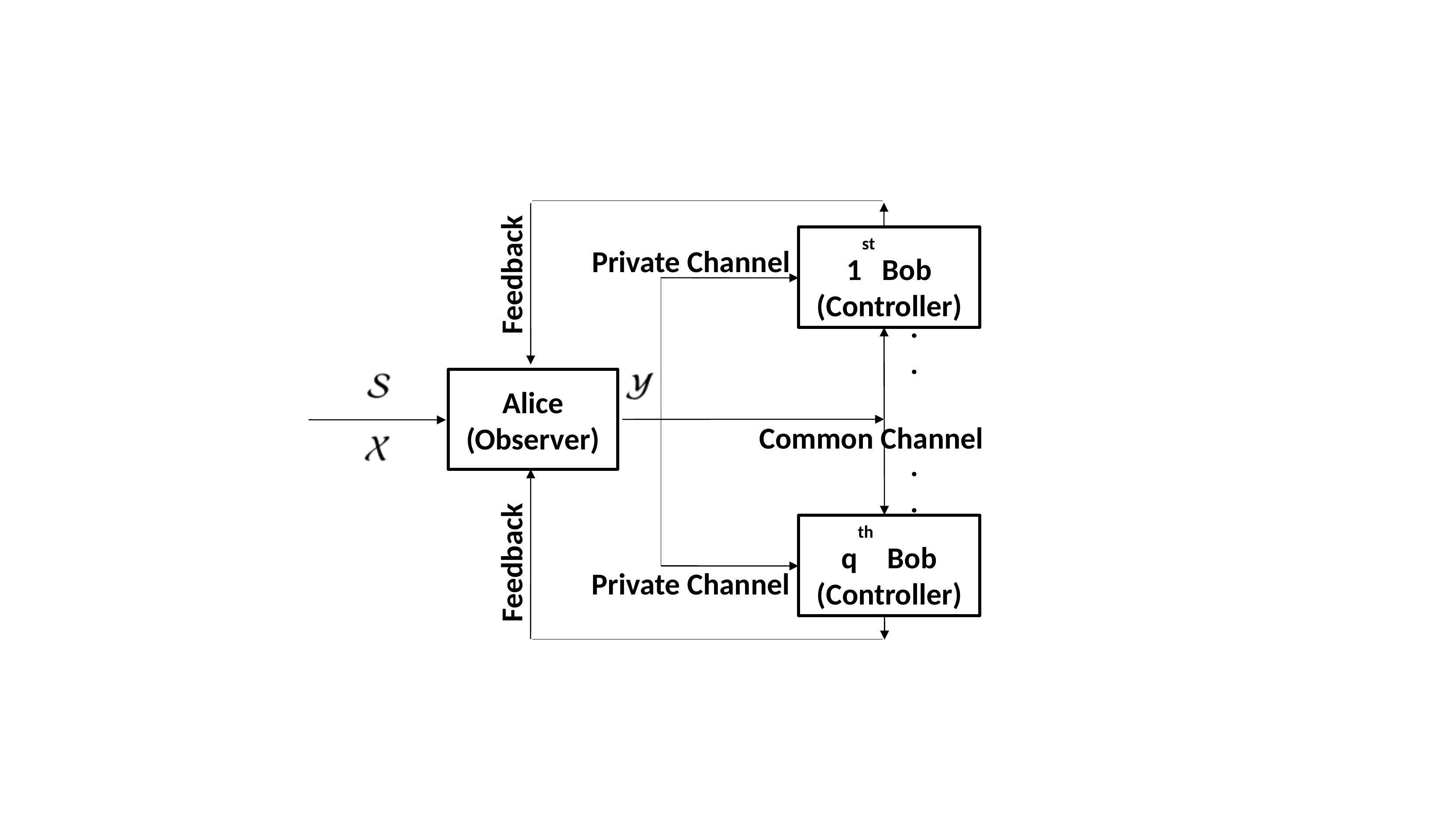}}\\
\caption{\textcolor{black}{A \textit{Gray-Wyner} system with generalised common information}.} \label{F1}
\label{fig:EcUND} 
\end{figure*}

\item We propose a novel \textit{mirror game} theoretical solution to \textcolor{black}{the \textit{Gray-Wyner} system with generalised common information}. We derive the appropriate conditions as well. \textcolor{black}{In order to tackle the possible risk of eavesdropping among Bobs, a virtual twin of the original message is created which is in the null of every Bob, but not in the null of others.}

\item We find an acceptably relaxed solution to the main problem while considering an uncertainty in our model. We do this from a worst-case optimisation method based information theoretic viewpoint. \textcolor{black}{We, in addition, make a comparison between our worst-case method based solution and a traditional non-convex optimisation approximation.}

\item We find a \textit{Nash} equilibrium for our mirror game theoretical solution with the use of \textit{contradiction}.

\item We also prove the inner stability of the closed-loop in our framework. 

\item We consider \textit{non-stationarity} for the control-input, consequently proposing three main solutions to deal with it. 
\begin{itemize}
\item First, a \textit{Lohe-model} based approach is proposed from a quantum-information-theoretic point of view. We do this with the goal of oscillator synchronisation analysis. 
\item  Additionally, a two-stage stochastic Stackelberg game is proposed which can optimally find the solution. In the later approach, we consider a non-stationary Major-player based scenario where the Leader solves the problem forward, whereas the Follower-set finds the solution backward in the given horizon. 
\item Thirdly, we formulate a novel mean-field-game (MFG) theoretical solution as well. In the last solution, the probability distribution function (PDF) of the MFG is also non-stationary for which we relax the problem. 
\end{itemize}
\item Finally, we explore the divergence and flux of the main problem from a generic point of view. The solutions expressed here are totally novel, and to the best of our knowledge, have not been explored so far.  
\end{itemize}

\subsection{General notation \& preliminaries} 

 Throughout the paper, the terms $\cdot \longmapsto \cdot$ and $ ( \cdot ||  \cdot    )$ stand for respectively the source-encoding/decoding process and Kullback-Leibler.

\textbf{\textsc{Assumption 1. \textit{Anytime-}capacity (\cite{2, 3, 5})}} \textit{Throughout the paper, when we say rate, we refer to the principle of anytime-capacity\footnote{The f$-$anytime capacity $C_{\alpha}(f)$ of a channel is the supremum of rates at which the channel can transmit data in the sense that (\textit{i}) the error probability is arbitrarily small, and (\textit{ii}) it decays at least as fast as $f(\cdot)$ does. $f(d) > 0$ is any decreasing function of the delay $d$ e.g. $f(d) = 2 ^{-\alpha d}$. Indeed, anytime capacity is the upper-bound of the error-free capacity being equated when $\alpha \rightarrow \infty$ and it is the lower-bound of the Shannon-capacity being equated when $\alpha \rightarrow 0$. The logic behind of the anytime capacity is the fact that the encoder-decoder pair must be anytime, i.e., timely synchronisable or real-time.}.}

\subsection{Organisation}
The rest of the paper is organised as follows. The system set-up and our main results \textcolor{black}{ $-$ including but not limited to the uncertainty-included and non-stationarity-included cases as well as the divergence and flux $-$ }are given in Sections II and III. Subsequently, the evaluation of the framework and conclusions are given in Sections IV and V. \textcolor{black}{The flow of the problem-and-solution is also depicted in Fig. \ref{F0}.}

\section{System model }
We here discuss about the system model, from both information theoretic and control theoretic standpoints. 

Consider Fig. \ref{F1} $-$ temporarily considering only one Bob. Assume that a sender named Alice has some private data denoted by the random variable $\mathcal{S} \in S$ which is correlated with some non-private data $\mathcal{X} \in X$. Alice is supposed to share $\mathcal{X}$ with an analyst named Bob. However, due to the correlation between $\mathcal{X}$ and $\mathcal{S}$ which is captured by the joint distribution $\mathcal{P}_{\mathcal{S},\mathcal{X}}$, Bob may be able to draw some inference on the private data $\mathcal{S}$. Alice consequently decides to, instead of $\mathcal{X}$, release a distorted version of $\mathcal{X}$ defined by $\mathcal{Y} \in Y$ in order to alleviate the inference threat over $\mathcal{S}$ acquirable from the observation of $\mathcal{X}$. The distorted data $\mathcal{Y}$ is generated by passing through the following privacy mapping, i.e., the conditional distribution $\mathcal{P}_{\mathcal{Y}|\mathcal{X}}$. It should be noted that, in fact, Bob may also be able to act as an adversary by using $\mathcal{Y}$ to illegitimately infer the private data set $\mathcal{S}$, even though he is a legitimate recipient of the data set $\mathcal{Y}$. Therefore, the privacy mapping should be designed in the sense that we can be assured about a reduction to the inference threat on the private set $\mathcal{S}$ as follows: while preserving the utility of $\mathcal{Y}$ by maintaining the correlation, i.e., dependency between $\mathcal{Y}$ and $\mathcal{X}$, we aim at alleviating the dependency between $\mathcal{S}$ and $\mathcal{Y}$. This kind of two-fold information-theoretic goal balances a trade-off between utility and privacy. As also obvious, the Markov chain $\mathcal{S} \rightarrow \mathcal{X}\rightarrow \mathcal{Y}$ holds. 

\subsection{From an information-theoretic point of view}
Let us go in datails from an information-theory point of view. For the $k$-the time instant where $k \in [0,\mathcal{K}]$ including $k \rightarrow \infty$ $-$ which declares that we use the \textit{any-time capacity} principle (see \textit{Assumption 1}), $-$ Alice observes the source-symbol sets $-$ sequences 
\begin{equation*}
\begin{split}
\mathcal{S}^{(\ell)}_k=\{ s_k^{(\ell)} \}^{\mathcal{L}_1}_{\ell=1},
\end{split}
\end{equation*}
and 
\begin{equation*}
\begin{split}
\mathcal{X}^{(\ell, q)}_k=\{ x_k^{(\ell, q)} \}^{\mathcal{L}_2}_{\ell=1},
\end{split}
\end{equation*}
where $q-$th Bob observes 
\begin{equation*}
\begin{split}
\mathcal{Y}^{(\ell, q)}_{k}=\{ y_{k}^{(\ell, q)} \}^{\mathcal{L}_3}_{\ell=1},
\end{split}
\end{equation*}
which are i.i.d discrete memoryless variables respectively with the probability mass functions $\mathcal{P}_{\mathcal{S}}$, $\mathcal{P}_{\mathcal{X}}$ and $\mathcal{P}_{\mathcal{Y}}$ where $\{ \cdot \}^{\big(\ell \in \left \{  1,\cdots,\mathcal{L}_i, i \in \{ 1,2,3 \}  \right \}  \big)}$ stands literally for the $\ell$-th source-symbol per block belonging to the source-symbol set of size $1$-by-$\mathcal{L}_i, i \in \{ 1,2,3 \}$. In fact, we principally see a simple scheme where: (\textit{i}) the Alice-encoder follows the Borel measurable map 
\begin{equation*}
\begin{split}
f^{(\ell, q)}_{x,k}: \;\mathcal{X}^{(\ell, q)}_{k}  \longmapsto \mathcal{M}^{(x,k, q)}=\{ 1, 2,\cdots, 2^{\ell \mathcal{R}^{(\ell, q)}_{x,k}} \},
\end{split}
\end{equation*}
while $\sum{q} \mathcal{R}^{(\ell, q)}_{x,k}$ is the rate of Alice; moreover, (\textit{ii}) the Bob-decoder legitimately follows the Borel measurable map 
\begin{equation*}
\begin{split}
g^{(\ell, q)}_{k}: \; \{ 1,\cdots, 2^{\ell \mathcal{R}^{(\ell, q)}_{x,k}} \}  \longmapsto \mathcal{X}^{(\ell, q)}_{k},
\end{split}
\end{equation*}
after reception via the channel \cite{22}.

\subsection{From a control-theoretic point of view}
Let us continue our discussion from a control-theory point of view. $\hat{\mathcal{X}}^{(\ell,q)}_{k}$ is the reconstructed version of $\mathcal{X}^{(\ell,q)}_{k} $ done by the $q-$th Bob-decoder, according to which he, as the controller, realises a map of $\hat{\mathcal{X}}^{(\ell,q)}_{k} \longmapsto \theta^{(\ell,q)}_{k}$. Now, we have a discrete-time stochastic linear system \cite{2, 3, 5} 
\begin{equation*}
\begin{split}
\dot{\mathscr{X}}_{k}=\mathscr{A}_1(k)\mathscr{X}(k)+\mathscr{A}_2(k)\mathscr{U}(k)+\dot{\mathscr{N}}_1(k),
\end{split}
\end{equation*}
where $\mathscr{X}(k) \in \mathbb{R}_n$ is the state, $\mathscr{N}_1(k) \in \mathbb{R}_n$ is the process noise, $\mathscr{U}(k) \in \mathbb{R}_m$ is the control action set, and $\mathscr{A}_1(k)$ and $\mathscr{A}_2(k)$ are matrices of sizes $n \times n$ and $n \times m$, respectively. At the time zone $k$, the controller observes the output $\mathscr{G}(k)$ of the channel, and chooses a control action $\mathscr{U}_{k}$ based upon the data it has inferred up to the time $k$. At time $k$, the encoder observes the output of the sensor $\mathscr{Y}(k) \in \mathbb{R}_j$ as 
\begin{equation*}
\begin{split}
\mathscr{Y}(k) = \mathscr{A}_3(k)\mathscr{X}(k) + \mathscr{N}_{2}(k),
\end{split}
\end{equation*}
where $\mathscr{A}_3(k)$ is a $j \times n$ deterministic matrix, and $\mathscr{N}_{2}(k) \in \mathbb{R}_j $ is the observation noise according to which $\mathscr{U}(k)=\mathscr{A}_4(k)\mathscr{Y}(k) $.

\section{Main results}
Main results are given in this section. We primarily introduce our proposed mirror game theoretical solution.

\subsection{General point of view}

\textsc{\textbf{Question 1:}} \textit{Can we interpret the generalised Gray-Wyner model from a mirror-game-theoretical point of view?} 
\textsc{\textbf{Answer:}} Our response is an affirmative one which is provided in the following \textcolor{black}{in-depth}. 

\begin{proposition} \label{P2} \textit{One can consider a mirror game scenario for the generalised Gray-Wyner model where Alice generates a vector of virtual messages as the virtual twins for the real message, i.e., $\big \lbrace\mathcal{Y}^{(tot)}_q\big \rbrace:=\Big \lbrace\mathcal{Y}^{(o)}_q;\big \lbrace\mathcal{Y}^{(v)}_{q}\big \rbrace \Big \rbrace$, where $\mathcal{Y}^{(o)}_q$ and $\big \lbrace\mathcal{Y}^{(v)}_{q}\big \rbrace$ respectively stand for the original and virtual data sets. Subsequently she sends it to each Bob where the virtual vector is in the null of the specific Bob, but not for the remaining ones.}\end{proposition} %


\textbf{\textsc{Proof:}} See Appendix \ref{sec:A}.$\; \; \; \blacksquare$

\subsection{Uncertainty included}

\begin{proposition} \label{P2} \textit{Our mirror-game-theoretical problem, although it is $NP-$hard under the assumption of experiencing uncertainties, it is relaxable.}\end{proposition} 

\textbf{\textsc{Proof:}} See Appendix \ref{sec:B}.$\; \; \; \blacksquare$

\textsc{\textit{Corollary 1:}} \textit{How to send the virtual twins, i.e., the virtual users in the null of the relative Bobs is performed in a relaxable fashion.}

\textsc{\textbf{Proof.}} See the Proposition 2 for the proof.$\; \; \; \blacksquare$

\begin{proposition} \label{P2} \textit{Our mirror-game-theoretical perspective undoubtedly experiences a \textit{Nash} equilibrium.}\end{proposition}

\textbf{\textsc{Proof:}} See Appendix \ref{sec:C}.$\; \; \; \blacksquare$

\begin{proposition} \label{P2} \textit{One can can construct encoders and controllers such that the closed loop system is stable and detectable for the tuple $(\mathscr{A}_1, \mathscr{A}_2, \mathscr{A}_3)$, i.e., the pair $(\mathscr{A}_1, \mathscr{A}_2)$ is controllable-and-stabilisable and the pair $(\mathscr{A}_1, \mathscr{A}_3)$ is observable-and-detectable.}\end{proposition}

\textbf{\textsc{Proof:}} See Appendix \ref{sec:D}.$\; \; \; \blacksquare$

\begin{figure}[t]
\centering
\subfloat{\includegraphics[trim={{10 mm} {64 mm} {21 mm} {74mm}},clip,scale=0.45]{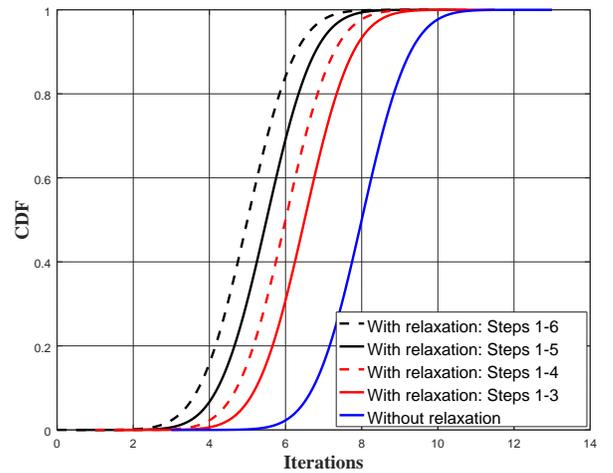}} 
\caption{CDF of the iterations required for the convergence of Algorithm \ref{fff} with and without the proposed relaxation. } \label{F2}
\label{fig:EcUND} 
\end{figure}

\begin{figure*}[t]
\centering
\subfloat[$|\mathscr{B}|=50 \%$]{\includegraphics[trim={{10 mm} {64 mm} {21 mm} {74mm}},clip,scale=0.45]{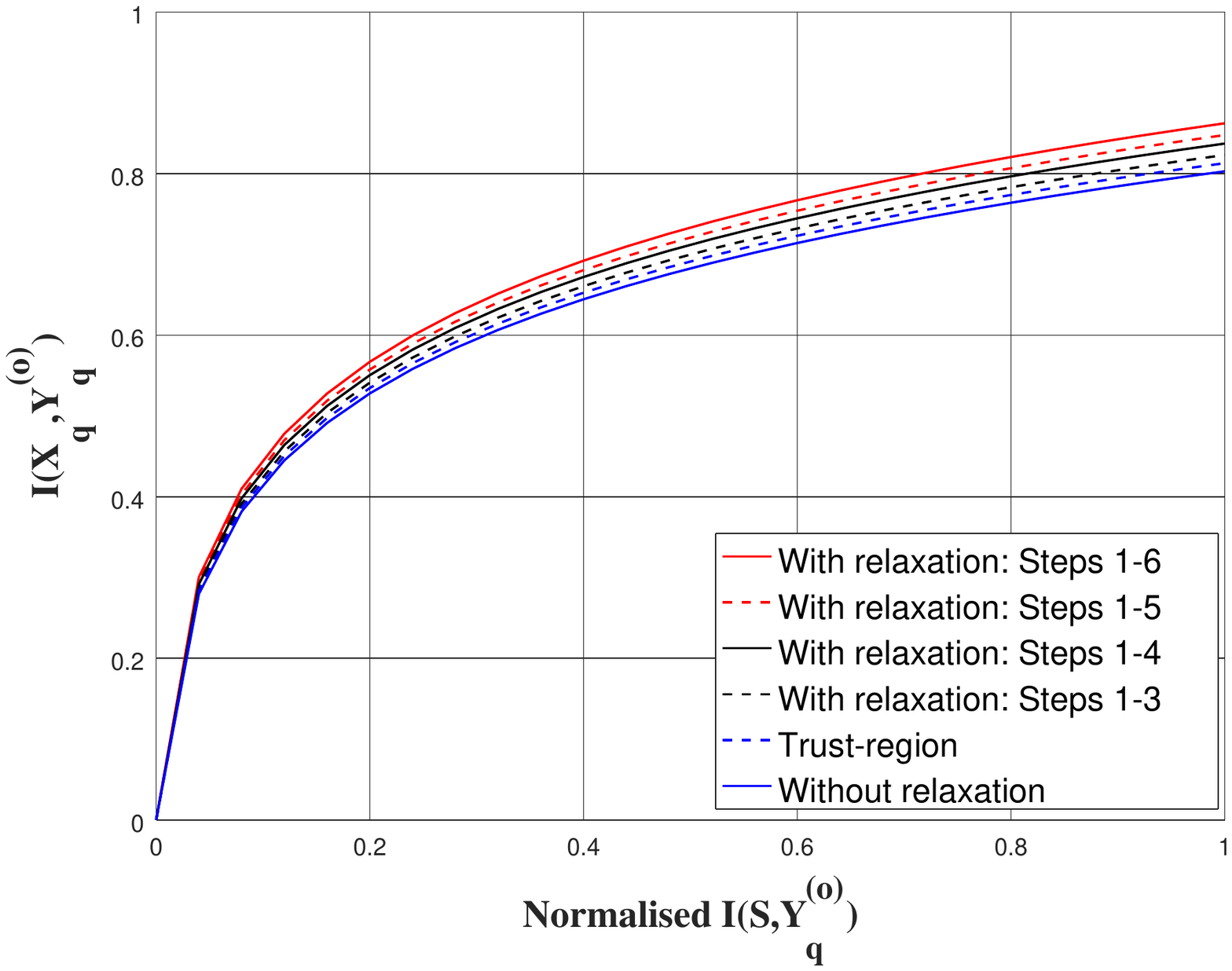}} 
\subfloat[$|\mathscr{B}|=10 \%$]{\includegraphics[trim={{10 mm} {64 mm} {21 mm} {74mm}},clip,scale=0.45]{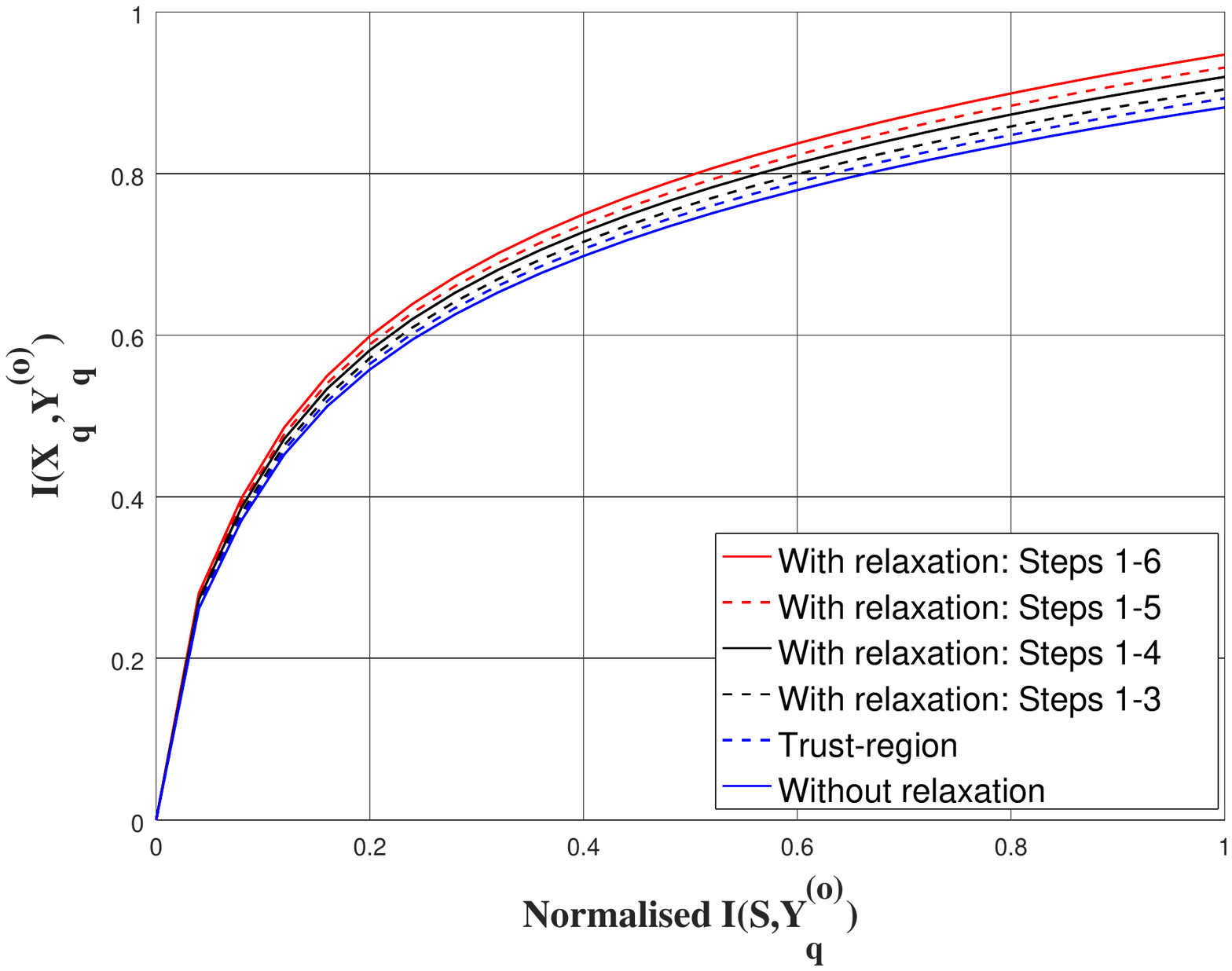}} 
\caption{$\mathcal{I} \big(\mathcal{X}_q,\mathcal{Y}^{(o)}_q\big)$ versus the normalised regime of $\mathcal{I} \big(\mathcal{S},\mathcal{Y}^{(o)}_q\big)$. } \label{F3}
\label{fig:EcUND} 
\end{figure*}

\subsection{Non-stationarity included}

First, let us define the control-input $\mathscr{U}(k)$ in a non-stationary fashion as
\begin{equation*}
\begin{split}
\mu_{\mathscr{U}(k+1)}=\mu_{\mathscr{U}(k)}+\xi(k),
\end{split}
\end{equation*}
which expresses that the mean of $\mathscr{U}(k)$ is non-stationary while $\xi(k)$ is a random-walk Wyner process.

\begin{lemma} \label{L1} \textit{One can re-write $\mathscr{U}(k)$ as $	\overrightarrow{\rm \mathscr{U}}(k)=\mathscr{U}_{{\mathcal{U}}_0(k)}	\overrightarrow{\rm {\mathcal{U}}_0(k)}+\mathscr{U}_{\hat{\mathcal{X}}(k)}	\overrightarrow{\rm \hat{\mathcal{X}}(k)}+\mathscr{U}_{{\mathcal{P}}(k)}	\overrightarrow{\rm \mathcal{P}(k)}$ in the context of the orthonormal basis vectors, regarding the normalised vectors $\overrightarrow{\rm {\mathcal{U}}_0(k)}$, $\overrightarrow{\rm \hat{\mathcal{X}}(k)}$ and $\overrightarrow{\rm \mathcal{P}(k)}$ $-$ a.k.a with $\mathcal{P}\big(\mathcal{Y}|\mathcal{X} \big)$ $-$ as well as the initial state $\mathcal{U}_0(k)$.}\end{lemma}
 
\textbf{\textsc{Proof:}} See Appendix \ref{sec:E}.$\; \; \; \blacksquare$

\subsubsection{Solution 1. Oscillator synchronisation analytical method}

\begin{proposition} \label{P2} \textit{One can consider the effect of the non-stationarity on dynamicity and dynamical charactristics of the network by modeling Oscillations.}\end{proposition}

\textbf{\textsc{Proof:}} See Appendix \ref{sec:F}.$\; \; \; \blacksquare$

\subsubsection{Solution 2. Non-stationary Major-Player Stackelberg game theoretical standpoint}

\begin{proposition} \label{P3} \textit{One can solve the main problem from a non-stationary Major-Player game theoretical point of view. }\end{proposition}

\textbf{\textsc{Proof:}} See Appendix \ref{sec:G}.$\; \; \; \blacksquare$

\subsubsection{Solution 3. Non-stationary Major-Player MFG theoretical standpoint}
\begin{proposition} \label{P1} \textit{One can find a MFG theoretical point of view in which the pdf of the major-player should be considered non-stationary.}\end{proposition}

\textbf{\textsc{Proof:}} See Appendix \ref{sec:H}.$\; \; \; \blacksquare$

\begin{proposition} \label{P1} \textit{Our MFG is of a smooth nature, thus, we can say that our system model is acceptably controllable-and-detectable.}\end{proposition}

\textbf{\textsc{Proof:}} See Appendix \ref{sec:I}.$\; \; \; \blacksquare$

\subsection{Divergence and flux from a generic perspective}

\begin{figure*}[t]
\centering
\subfloat[$|\mathscr{B}|=60 \%$]{\includegraphics[trim={{10 mm} {64 mm} {21 mm} {74mm}},clip,scale=0.45]{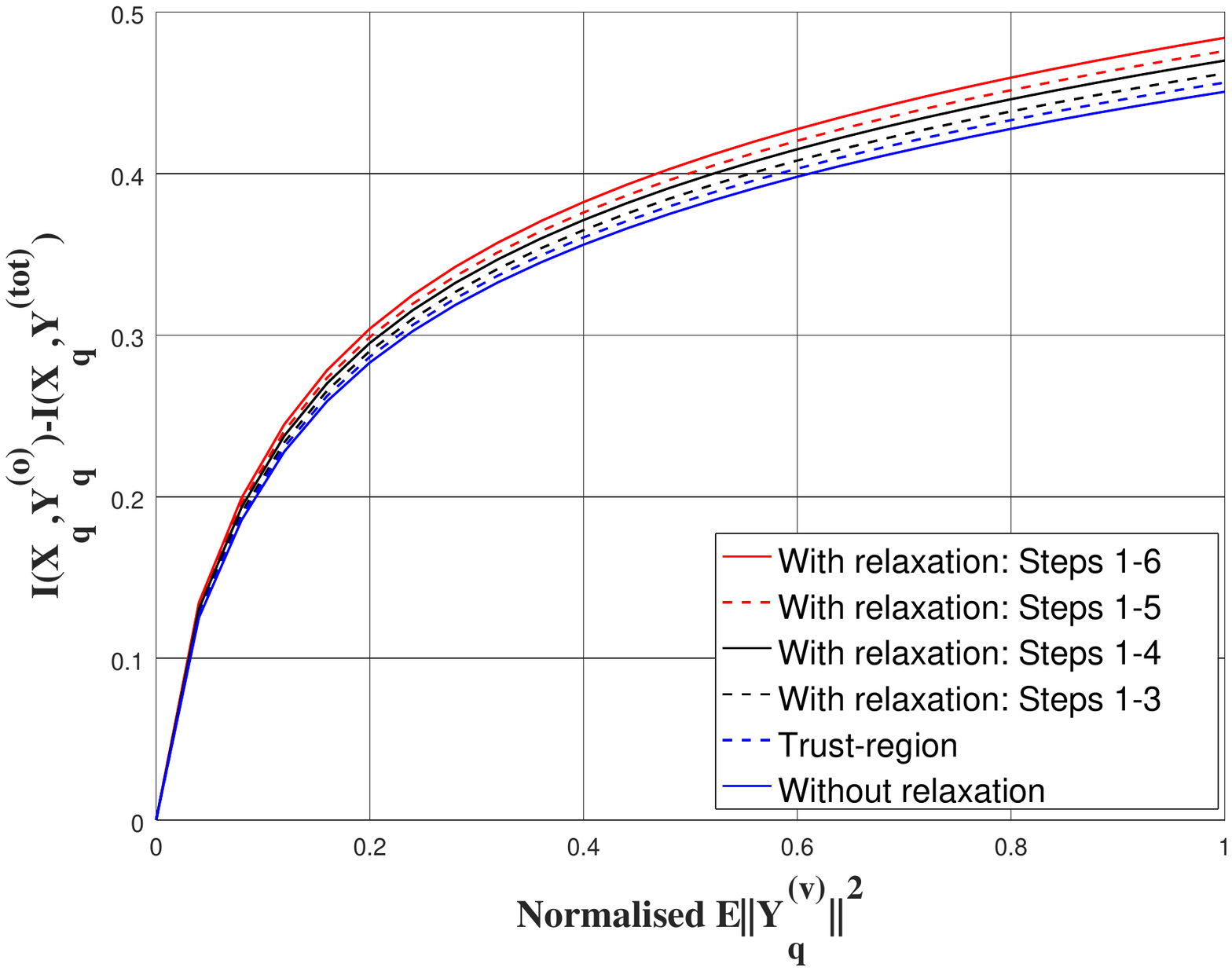}} 
\subfloat[$|\mathscr{B}|=70 \%$]{\includegraphics[trim={{10 mm} {64 mm} {21 mm} {74mm}},clip,scale=0.45]{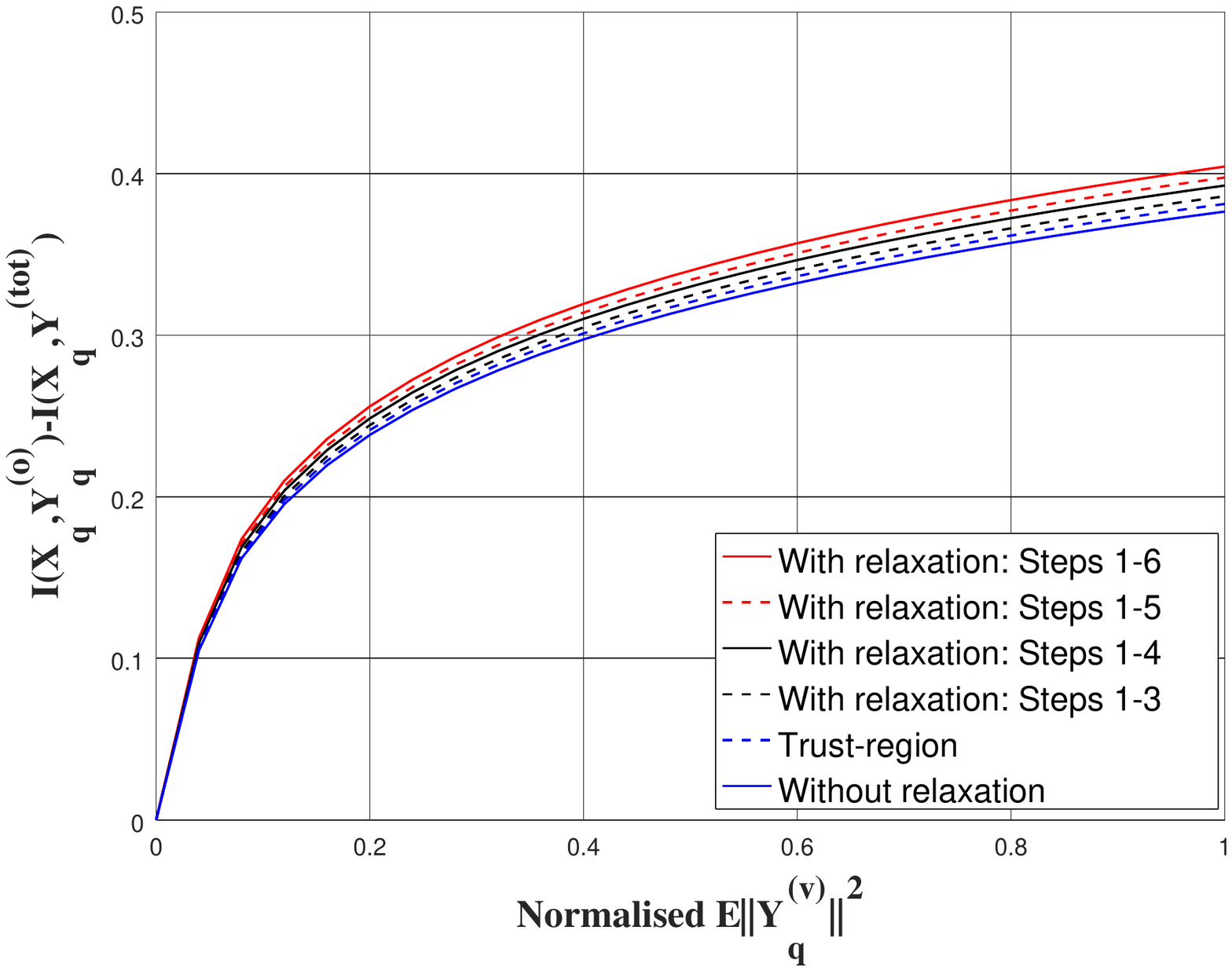}} 
\caption{$\mathcal{I} \big(\mathcal{X}_q,\mathcal{Y}^{(o)}_q\big)- \mathcal{I} \big(\mathcal{X}_q,\big \lbrace\mathcal{Y}^{(tot)}_{q^{\prime}}\big \rbrace\big)$ versus the normalised regime of $\mathbb{E} \big \lbrace || \big \lbrace\mathcal{Y}^{(v)}_{q}\big \rbrace||^2\big \rbrace \le \gamma^{(q)}_1$.} \label{F4}
\label{fig:EcUND} 
\end{figure*}

\begin{proposition} \label{P1} \textit{Calling $\mathcal{Z}$ for the \textcolor{black}{common information} from a generic standpoint, the divergence of $\mathop{{\rm max}}\limits_{(\cdot)} {\rm \; } \mathcal{I}\big(\mathcal{X},\mathcal{Y}|\mathcal{Z}\big)$ is acceptably followable. }\end{proposition}
 
\textbf{\textsc{Proof:}} See Appendix \ref{sec:J}.$\; \; \; \blacksquare$

\begin{proposition} \label{P1} \textit{In the subsequence of Proposition 9 and in the case of uncertainties, the divergence named above can be acceptably followable.  }\end{proposition}
 
\textbf{\textsc{Proof:}} See Appendix \ref{sec:K}.$\; \; \; \blacksquare$

\section{Numerical results}

We have done our simulations w.r.t. the Bernoulli-distributed data-sets using GNU Octave of version $4.2.2$
on Ubuntu $16.04$. 

Initially speaking, a greedy algorithm as \ref{fff} is exemplified here which can solve our main problem. Meanwhile, Algorithm \ref{ffff} is presented in terms of the \textit{trust-region} method\footnote{See e.g. \cite{23} in order to understand what it is.} based non-convex programming which is used in Figs. \ref{F3} and \ref{F4}.

\begin{algorithm*}
\caption{\textcolor{black}{A greedy algorithm to the Problem $ \mathscr{P}_1$.}}\label{fff}
\begin{algorithmic}
\STATE \textbf{\textsc{Initialisation.}} 

\textbf{while} $ \mathbb{TRUE}$ \textbf{do} 

$\;\;\;\;\;\;\;\;\;\;\;\;\;\;\;\;\;\;\;\;\;\;\;\;$(\textit{i}) Solve $\mathscr{P}_1$; and 

$\;\;\;\;\;\;\;\;\;\;\;\;\;\;\;\;\;\;\;\;\;\;\;\;$(\textit{ii}) Update. 

\textbf{endwhile} 

\textbf{\textsc{Output.}} 

\textbf{end}

\end{algorithmic}
\end{algorithm*}

\begin{algorithm*}
\caption{\textcolor{black}{Trust-region method to solve the Problem $ \mathscr{P}_1$.}}\label{ffff}
\begin{algorithmic}
\STATE \textbf{\textsc{Initialisation:}} $0 < \eta_1 < \eta_2 < 1; 0 < \theta_1 < \theta_2 ;$ and $x_0$.

$\;\;$\textbf{while} $ ||\nabla_{\xi}\psi(\xi_n) \ge \epsilon_{th}||$ \textbf{do} 

$\;\;\;\;$ (\textit{i}) Calculate the trial step by: $\zeta_{m}:=arg\mathop{{\rm min}}\limits_{|\zeta|\le\Delta_m} {\rm \; } \psi_m (\zeta)+\nabla_{\xi}\psi(\xi_m)^{T} \zeta+\frac{1}{2}\zeta^{T}\nabla^{2}_{\xi}\psi(\xi_m)\zeta$

$\;\;\;$ (\textit{ii}) Test-and-update the trial step and the trust-region radius $\Delta_m$ while $\tau_m:=\frac{\psi(\xi_m+\zeta_m)-\psi(\xi_m)}{\psi_m(\zeta)-\psi_m(0)}$:

$\;\;$ $\;\;$ $\;\;$ \textbf{if} $\tau_m>\eta_1, \xi_{m+1} \gets \xi_m+\zeta_m$

$\;\;$ $\;\;$ $\;\;$ \textbf{else} $\xi_{m+1} \gets \xi_m$, \textbf{endif} 

$\;\;$ $\;\;$ $\;\;$ \textbf{if} $\tau_m \le \eta_1, \Delta_{m+1} \gets \theta_1 ||\zeta_m|| $

$\;\;$ $\;\;$ $\;\;$ \textbf{elseif} $\tau_m > \eta_2 $ and $||\zeta_m||=\Delta_m, \Delta_{m+1} \gets \theta_2 \Delta_m$

$\;\;$ $\;\;$ $\;\;$ \textbf{else} $\Delta_{m+1} \gets \Delta_m$, \textbf{endif} 

$\;\;$\textbf{endwhile} 

$m \gets m+1$

\textbf{end}

\end{algorithmic}
\end{algorithm*}

Fig. \ref{F2} shows the cumulative distribution function (CDF) of the iterations needed for the Algorithm \ref{fff} $-$ applicable to solve the Problem $\mathscr{P}_1$. As obvious, Algorithm \ref{fff} performs better by applying our proposed relaxation interpretation. More interestingly, this figure completely proves that the more our proposed relaxation method is applied, the more performance we can experience. 

Fig. \ref{F3} shows $\mathcal{I} \big(\mathcal{X}_q,\mathcal{Y}^{(o)}_q\big)$ versus the normalised regime of $\mathcal{I} \big(\mathcal{S},\mathcal{Y}^{(o)}_q\big)$ while changing the amount of the uncertainty $\mathscr{B}$. The effect of our proposed relaxation method is totally obvious the same as the previous figure. 

Fig. \ref{F4} shows $\mathcal{I} \big(\mathcal{X}_q,\mathcal{Y}^{(o)}_q\big)- \mathcal{I} \big(\mathcal{X}_q,\big \lbrace\mathcal{Y}^{(tot)}_{q^{\prime}}\big \rbrace\big)$ versus the normalised regime of $\mathbb{E} \big \lbrace || \big \lbrace\mathcal{Y}^{(v)}_{q}\big \rbrace||^2\big \rbrace $ while changing the amount of the uncertainty $\mathscr{B}$. Our proposed relaxation method still leads. Indeed, this figure analyses the secrecy rate for our mirror game theoretical scheme. Moreover, the value for $q^{\prime}$ is assumed on average in this figure, that is, we ignored the term $\frac{1}{q^{\prime}}\sum{q^{\prime}}$ for the ease of notation since we talk about the average rate per user. 

\begin{figure*}[t]
\centering
\subfloat[Iterations]{\includegraphics[trim={{17 mm} {64 mm} {21 mm} {74mm}},clip,scale=0.44]{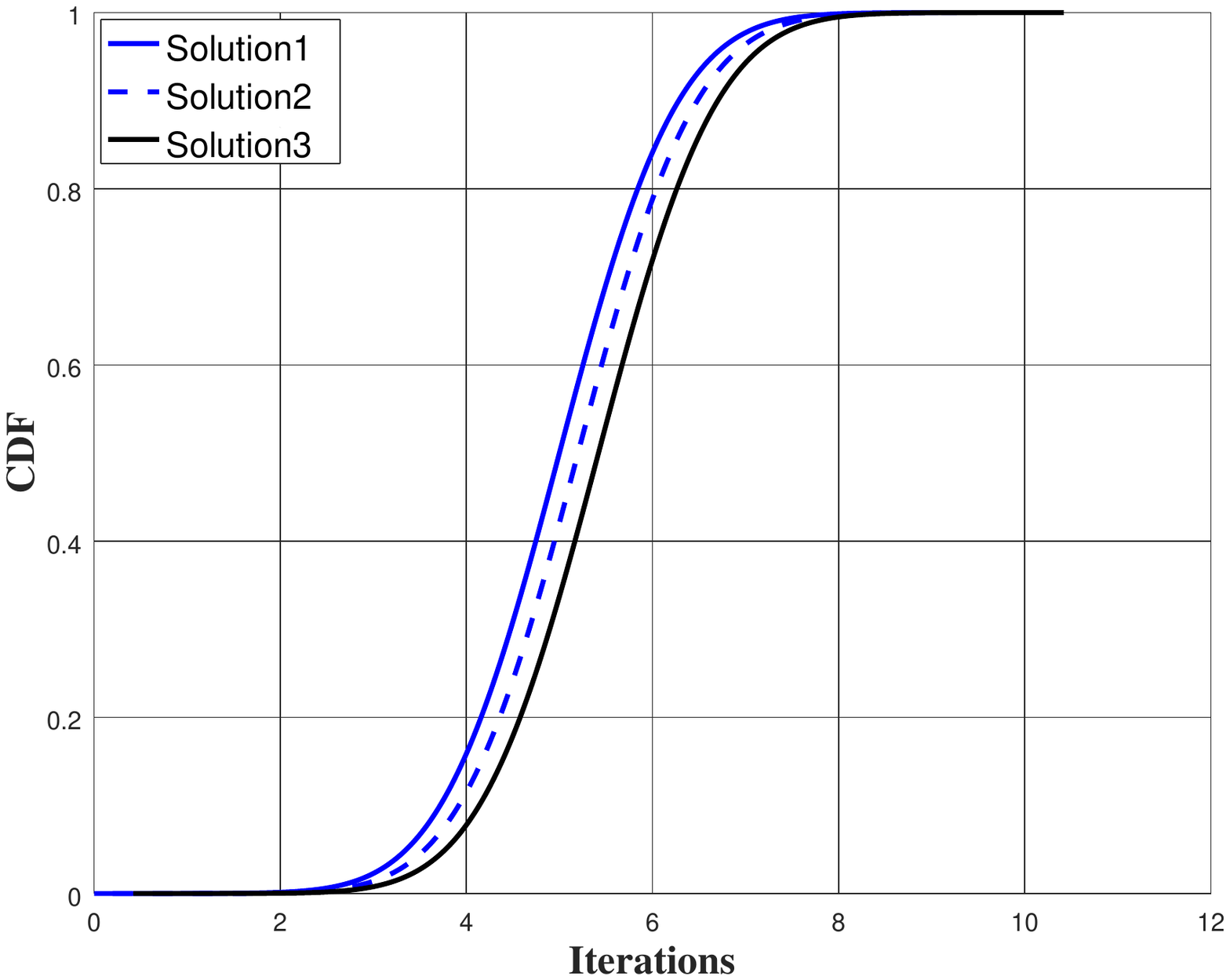}} 
\subfloat[Accuracy]{\includegraphics[trim={{17 mm} {64 mm} {21 mm} {74mm}},clip,scale=0.44]{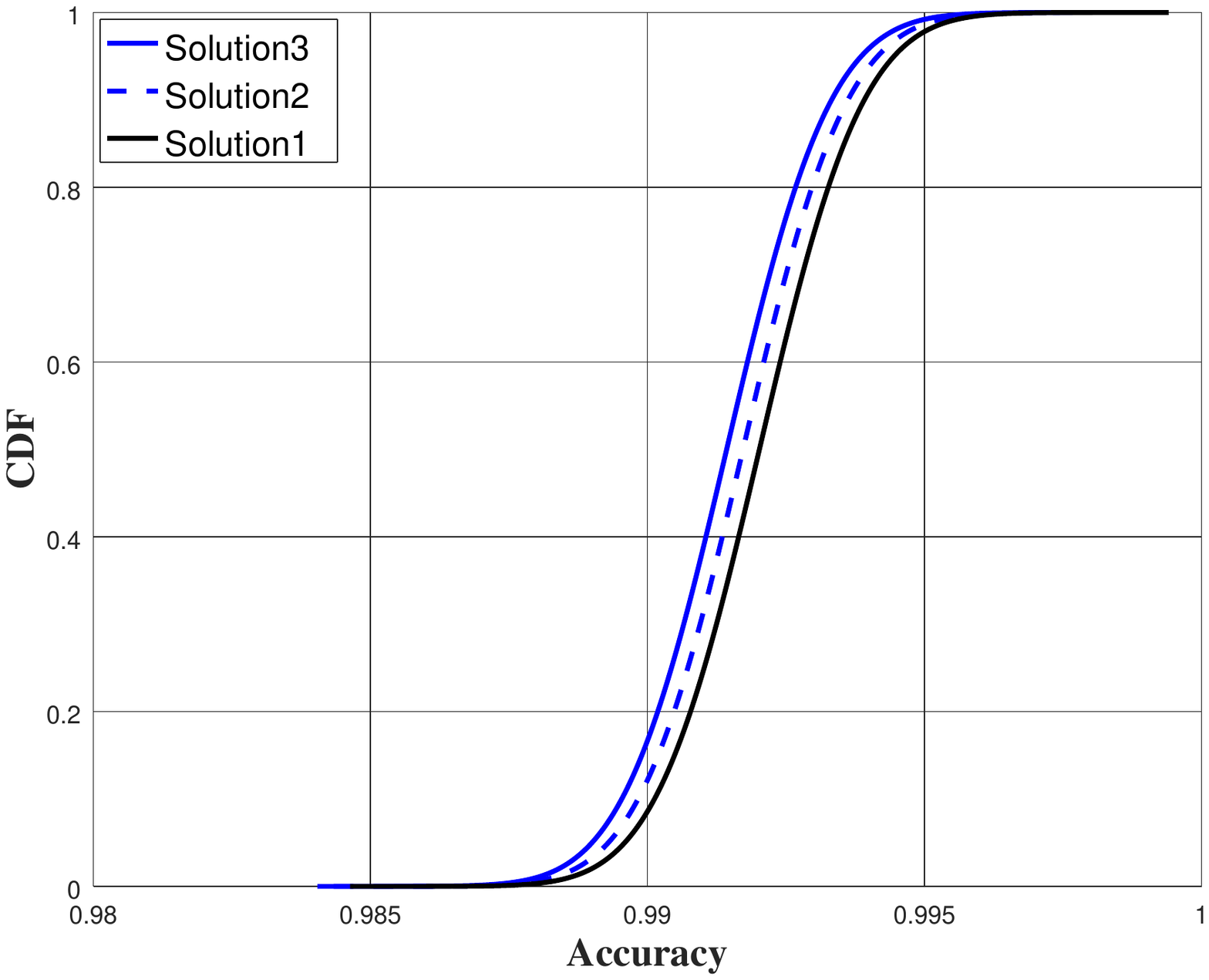}} 
\caption{Non-stationarity included case: CDFs for a given greedy algorithm. } \label{F5}
\label{fig:EcUND} 
\end{figure*}

Fig. \ref{F5} demonstrates, for the case of non-stationarity included, the CDF of respectively: (\textit{i}) the iterations needed for a given greedy algorithm to be converged; and (\textit{ii}) the accuracy. As obvious, the performances are approximately near while the solutions $1$ and $3$ respectively have the slightly best and the worst performances.

\section{conclusion}
A novel \textit{mirror game} theoretical solution was proposed to the \textcolor{black}{\textit{Gray-Wyner} schemes with generalised common information}. A relaxed solution to the main problem was proven while considering an uncertainty in the paradigm in terms of the worst-case method. The existence of a \textit{Nash} equilibrium for our mirror game was also proven by \textit{contradiction}. We also proved the inner stability of the closed-loop in our model. Meanwhile, the case of non-stationarity included was explored proposing three solutions. Simulations finally proved that our relaxation method is extremely useful, particularly, every step in-hand.

\appendices
\section{Proof of Proposition 1}
\label{sec:A}
Our proposed strategy should be logically able to simultaneously guarantee the following conditions. We define seven ones as:
\begin{itemize}
\item (\textit{i}) 
\begin{equation*}
\begin{split}
\mathop{{\rm max}}\limits_{\mathcal{P}{\big(\mathcal{Y}^{(o)}_q|\mathcal{X}_q} \big)} {\rm \; } \mathcal{I} \big(\mathcal{X}_q,\mathcal{Y}^{(o)}_q\big),
\end{split}
\end{equation*}
\textcolor{black}{as the \textit{rate-distortion} theory emphatically entails}; 
\item (\textit{ii}) 
\begin{equation*}
\begin{split}
\mathcal{I} \big(   \mathcal{Y}^{(o)}_q,\mathcal{S} \big) \le \gamma^{(q)}_0,
\end{split}
\end{equation*}
\textcolor{black}{as the \textit{rate-distortion} theory emphatically entails}; 
\item (\textit{iii}) 
\begin{equation*}
\begin{split}
\mathop{{\rm min}}\limits_{\mathcal{P}{\big(\big \lbrace\mathcal{Y}^{(tot)}_{q^{\prime}}\big \rbrace|\mathcal{X}_q} \big)} {\rm \; } \mathcal{I} \big(\mathcal{X}_q,\big \lbrace\mathcal{Y}^{(tot)}_{q^{\prime}}\big \rbrace\big), q^{\prime} \neq q,
\end{split}
\end{equation*}
\textcolor{black}{in order to guarantee the privacy among the users}; 
\item (\textit{iv}) 
\begin{equation*}
\begin{split}
\mathbb{E} \big \lbrace ||   \big \lbrace\mathcal{Y}^{(v)}_{q}\big \rbrace||^2\big \rbrace \le \gamma^{(q)}_1,
\end{split}
\end{equation*}
\textcolor{black}{as our design resources are limited}; 
\item (\textit{v})\footnote{\textcolor{black}{Since one is related to the $q$th Bob and another one is related to the $q^{\prime}$th Bob.}} 
\begin{equation*}
\begin{split}
\mathcal{I} \big(   \big \lbrace\mathcal{Y}^{(v)}_{q^{\prime}}\big \rbrace,\mathcal{Y}^{(o)}_{q} \big)>0,
\end{split}
\end{equation*}
\textcolor{black}{in order to guarantee the privacy among the users}; 
\item (\textit{vi})\footnote{\textcolor{black}{Since one is related to the $q$th Bob and another one is related to the $q^{\prime}$th Bob.}} 
\begin{equation*}
\begin{split}
\mathcal{I} \big( \big \lbrace  \mathcal{Y}^{(v)}_{q^{\prime}}\big \rbrace,\mathcal{X}_{q} \big)>0,
\end{split}
\end{equation*}
\textcolor{black}{in order to guarantee the privacy among the users}; and 
\item (\textit{vii}) 
\begin{equation*}
\begin{split}
\mathcal{I} \big( \big \lbrace  \mathcal{Y}^{(v)}_{q}\big \rbrace,\mathcal{Y}^{(o)}_{q} \big)=0,
\end{split}
\end{equation*}
\textcolor{black}{in order to guarantee the fairness\footnote{\textcolor{black}{Quality-of-service.}} for each individual user}, with regard to the non-zero positive arbitary thresholds $\gamma_0$ and $\gamma_1$.
\end{itemize}

In relation to the seven conditions introduced above, we \textcolor{black}{re-}express the following discussion. The first and the second conditions guarantee the privacy funnel. In addition, the third condition actualises a consolidation against the possible risk of some Bobs being potential adversaries. Pen-ultimately, the forth condition imposes that the number of the virtual twins must be constrained. Finally, the last three conditions show how the virtual twins are in the null of the relative receiver, but not for the others. 

\textsc{\textbf{Remark 1.}} \textit{Some extra conditions could have been defined, however, they are similar to others and thus ignorable, such as: $\mathcal{I} \big(\mathcal{X}_q,\big \lbrace\mathcal{Y}^{(v)}_{q}\big \rbrace\big)=0$ or/and $\mathcal{I} \big(   \mathcal{Y}^{(o)}_{q^{\prime}},\mathcal{S} \big) \le \cdots$.}

This completes the proof.$\; \; \; \blacksquare$

\section{Proof of Proposition 2}
\label{sec:B}

The proof is given in terms of the following multi-step solution.

\textsc{Step 1.} Let us call $\mathscr{B}$ as the uncertainty as well as providing the definition of another non-zero positive arbitary threshold $\gamma_2$. One can integrate and re-write the seven conditions derived above as the Problem $\mathscr{P}_1$ as \textcolor{black}{$-$ as \textit{rate-distortion} theory entails from a generic point of view}
\begin{equation*}
\begin{split}
(\mathscr{P}_1): \;\;\mathop{{\rm min}}\limits_{(\cdot)} {\rm \; } \mathcal{I} \big(\mathcal{X}_q(\mathscr{B}),\big \lbrace\mathcal{Y}^{(tot)}_{q^{\prime}}(\mathscr{B})\big \rbrace\big), q^{\prime} \neq q,\\
s.t. \;\; \mathcal{I} \big(\mathcal{X}_q(\mathscr{B}),\mathcal{Y}^{(o)}_q(\mathscr{B})\big) \ge \gamma_2,\;\;\;\;\;\;\;\;\;\\
\mathcal{I} \big( \mathcal{Y}^{(o)}_q(\mathscr{B}),\mathcal{S}(\mathscr{B})\big) \le \gamma^{(q)}_0,\;\;\;\;\;\;\;\;\\
\mathbb{E} \big \lbrace || \big \lbrace\mathcal{Y}^{(v)}_{q}(\mathscr{B})\big \rbrace||^2\big \rbrace \le \gamma^{(q)}_1,\;\;\;\;\;\;\;\;\;\\
\mathcal{I} \big( \big \lbrace\mathcal{Y}^{(v)}_{q^{\prime}}(\mathscr{B})\big \rbrace,\mathcal{Y}^{(o)}_{q} (\mathscr{B})\big)>0,\;\;\;\;\;\\
\mathcal{I} \big( \big \lbrace \mathcal{Y}^{(v)}_{q^{\prime}}(\mathscr{B})\big \rbrace,\mathcal{X}_{q} (\mathscr{B})\big)>0,\;\;\;\;\;\;\;\;\\
\mathcal{I} \big( \big \lbrace \mathcal{Y}^{(v)}_{q}(\mathscr{B})\big \rbrace,\mathcal{Y}^{(o)}_{q} (\mathscr{B})\big)=0.\;\;\;\;\;\;
\end{split}
\end{equation*}

We know e.g. from \cite{24} that the \textit{worst-case} method allows us to recast 
\begin{equation*}
\begin{split}
\mathop{{\rm max}}\limits_{(x)} {\rm \; } t(x;\mathscr{B}),
\end{split}
\end{equation*}
into
\begin{equation*}
\begin{split}
\mathop{{\rm max}}\limits_{x} {\rm \; } t_1(x) \;\; s.t. \;\; t_1(x) \le t(x;\mathscr{B}).
\end{split}
\end{equation*}
$s$. \textcolor{black}{Therefore, our mission is now to examine if we can apply the worst-case method on $(\mathscr{P}_1)$.}

\textsc{Step 2.} In the consequence of the point expressed above, we intelligently re-define the optimisation problem derived above from a probabilistic point of view, where $q^{\prime} \neq q$ holds in parallel with the definition of another non-zero positive arbitary thresholds $\gamma_3$ and $\vartheta_i, i \in \{ 0, 1, 2, 3, 4, 5, 6\}$, as
\begin{equation*}
\begin{split}
\mathop{{\rm max}}\limits_{(\cdot)} {\rm \; } \vartheta_0\textcolor{black}{(\mathscr{B})}, \;\;\;\;\;\;\;\;\;\;\;\;\;\;\;\;\;\;\;\;\;\;\;\;\;\;\;\;\;\;\;\;\;\;\;\;\;\;\;\;\;\;\;\;\;\;\;\;\;\;\;\;\;\;\;\;\;\;\;\;\\
\mathscr{Pr} \Big \lbrace \mathcal{I} \big(\mathcal{X}_q(\mathscr{B}),\big \lbrace\mathcal{Y}^{(tot)}_{q^{\prime}}(\mathscr{B})\big \rbrace\big) \le \gamma_3 \Big \rbrace \ge \vartheta_0\textcolor{black}{(\mathscr{B})},\;\;\;\;\;\;\;\;\;\;\\
s.t. \;\; \mathscr{Pr} \Big \lbrace\mathcal{I} \big(\mathcal{X}_q(\mathscr{B}),\mathcal{Y}^{(o)}_q(\mathscr{B})\big) \ge \gamma_2 \Big \rbrace \ge \vartheta_1\textcolor{black}{(\mathscr{B})},\;\;\;\;\;\;\;\;\;\;\;\;\;\;\;\;\\
\mathscr{Pr} \Big \lbrace\mathcal{I} \big( \mathcal{Y}^{(o)}_q(\mathscr{B}),\mathcal{S}(\mathscr{B})\big) \le \gamma^{(q)}_0 \Big \rbrace \ge \vartheta_2\textcolor{black}{(\mathscr{B})},\;\;\;\;\;\;\;\;\;\;\;\;\;\;\;\\
\mathscr{Pr} \Big \lbrace\mathbb{E} \big \lbrace || \big \lbrace\mathcal{Y}^{(v)}_{q}(\mathscr{B})\big \rbrace||^2\big \rbrace \le \gamma^{(q)}_1 \Big \rbrace \ge \vartheta_3\textcolor{black}{(\mathscr{B})},\;\;\;\;\;\;\;\;\;\;\;\;\;\;\;\;\\
\mathscr{Pr} \Big \lbrace\mathcal{I} \big( \big \lbrace\mathcal{Y}^{(v)}_{q^{\prime}}(\mathscr{B})\big \rbrace,\mathcal{Y}^{(o)}_{q} (\mathscr{B})\big)>0 \Big \rbrace \ge \vartheta_4\textcolor{black}{(\mathscr{B})},\;\;\;\;\;\;\;\;\;\;\;\;\\
\mathscr{Pr} \Big \lbrace\mathcal{I} \big( \big \lbrace \mathcal{Y}^{(v)}_{q^{\prime}}(\mathscr{B})\big \rbrace,\mathcal{X}_{q} (\mathscr{B})\big)>0 \Big \rbrace \ge \vartheta_5\textcolor{black}{(\mathscr{B})},\;\;\;\;\;\;\;\;\;\;\;\;\;\;\;\\
\mathscr{Pr} \Big \lbrace\mathcal{I} \big( \big \lbrace \mathcal{Y}^{(v)}_{q}(\mathscr{B})\big \rbrace,\mathcal{Y}^{(o)}_{q} (\mathscr{B})\big)=0 \Big \rbrace \ge \vartheta_6(\textcolor{black}{\mathscr{B})},\;\;\;\;\;\;\;\;\;\;\;\;
\end{split}
\end{equation*}
or equivalently 
\begin{equation*}
\begin{split}
\mathop{{\rm max}}\limits_{(\cdot)} {\rm \; } \;\;\;\;\;\;\;\;\;\;\;\;\;\;\;\;\;\;\;\;\;\;\;\;\;\;\;\;\;\;\;\;\;\;\;\;\;\;\;\;\;\;\;\;\;\;\;\;\;\;\;\;\;\;\;\;\;\;\;\;\;\;\;\;\;\;\;\;\;\;\\ \big \lbrace \vartheta_0(\mathscr{B}), \vartheta_1(\mathscr{B}), \vartheta_2(\mathscr{B}), \vartheta_3(\mathscr{B}), \vartheta_4(\mathscr{B}), \vartheta_5(\mathscr{B}), \vartheta_6(\mathscr{B})\big \rbrace. 
\end{split}
\end{equation*}

\textsc{Step 3.} The last equation expressed above is still $NP-$ hard although it can be solved e.g. by the \textit{alternating direction method of multipliers} based \textit{simultaneous perturbation stochastic approximation} derived methods\footnote{See e.g. \cite{2}.}. Thus, we still undoubtedly need to continue. 


Let us carefully take a look at the above-mentioned equations. We see that 
\begin{equation*}
\begin{split}
\mathscr{Pr} \Big \lbrace\mathcal{I} \big( \big \lbrace\mathcal{Y}^{(v)}_{q^{\prime}}(\mathscr{B})\big \rbrace,\mathcal{Y}^{(o)}_{q} (\mathscr{B})\big)>0 \Big \rbrace,
\end{split}
\end{equation*}
\begin{equation*}
\begin{split}
\mathscr{Pr} \Big \lbrace\mathcal{I} \big( \big \lbrace \mathcal{Y}^{(v)}_{q^{\prime}}(\mathscr{B})\big \rbrace,\mathcal{X}_{q} (\mathscr{B})\big)>0 \Big \rbrace,
\end{split}
\end{equation*}
and
\begin{equation*}
\begin{split}
\mathscr{Pr} \Big \lbrace\mathcal{I} \big( \big \lbrace \mathcal{Y}^{(v)}_{q}(\mathscr{B})\big \rbrace,\mathcal{Y}^{(o)}_{q} (\mathscr{B})\big)=0 \Big \rbrace,
\end{split}
\end{equation*}
are more stretched compared to the other constraints since they are upper-bounded by $e^{0}$, according to the \textit{Concentration-of-Measure inequalities}. Now one can find the very small non-zero values $\epsilon_1^+$, $\epsilon_2^+$ and $\epsilon_3^+$ in the sense that 
\begin{equation*}
\begin{split}
\mathscr{Pr} \Big \lbrace\mathcal{I} \big( \big \lbrace\mathcal{Y}^{(v)}_{q^{\prime}}(\mathscr{B})\big \rbrace,\mathcal{Y}^{(o)}_{q} (\mathscr{B})\big)>\epsilon_1^+ \Big \rbrace \ge \vartheta^{\prime}_4,
\end{split}
\end{equation*}
\begin{equation*}
\begin{split}
\mathscr{Pr} \Big \lbrace\mathcal{I} \big( \big \lbrace \mathcal{Y}^{(v)}_{q^{\prime}}(\mathscr{B})\big \rbrace,\mathcal{X}_{q} (\mathscr{B})\big)>\epsilon_2^+ \Big \rbrace \ge \vartheta^{\prime}_5,
\end{split}
\end{equation*}
and
\begin{equation*}
\begin{split}
\mathscr{Pr} \Big \lbrace\mathcal{I} \big( \big \lbrace \mathcal{Y}^{(v)}_{q}(\mathscr{B})\big \rbrace,\mathcal{Y}^{(o)}_{q} (\mathscr{B})\big) \ge \epsilon_3^+ \Big \rbrace \ge \vartheta^{\prime}_6,
\end{split}
\end{equation*}
are still valid \textcolor{black}{$-$ for the given arbitrary thresholds $ \vartheta^{\prime}_4,  \vartheta^{\prime}_5,  \vartheta^{\prime}_6$ which are not functions of $\mathscr{B}$ anymore, as the worst-case method entails $-$} while the three later expressions can be the lower-bounds of respectively 
\begin{equation*}
\begin{split}
\mathscr{Pr} \Big \lbrace\mathcal{I} \big( \big \lbrace\mathcal{Y}^{(v)}_{q^{\prime}}(\mathscr{B})\big \rbrace,\mathcal{Y}^{(o)}_{q} (\mathscr{B})\big)>0 \Big \rbrace \ge \vartheta_4\textcolor{black}{(\mathscr{B})},
\end{split}
\end{equation*}
\begin{equation*}
\begin{split}
\mathscr{Pr} \Big \lbrace\mathcal{I} \big( \big \lbrace \mathcal{Y}^{(v)}_{q^{\prime}}(\mathscr{B})\big \rbrace,\mathcal{X}_{q} (\mathscr{B})\big)>0 \Big \rbrace \ge \vartheta_5\textcolor{black}{(\mathscr{B})},
\end{split}
\end{equation*}
and
\begin{equation*}
\begin{split}
\mathscr{Pr} \Big \lbrace\mathcal{I} \big( \big \lbrace \mathcal{Y}^{(v)}_{q}(\mathscr{B})\big \rbrace,\mathcal{Y}^{(o)}_{q} (\mathscr{B})\big) =0 \Big \rbrace \ge \vartheta_6\textcolor{black}{(\mathscr{B})},
\end{split}
\end{equation*}
consequently, these three aforementioned bounds can be relaxed by the \textit{Concentration-of-Measure inequalities}\footnote{\textcolor{black}{See e.g. \cite{b, bb}.}}. Therefore, the optimisation problem derived above, $-$ although it has not been completely relaxed $-$ , can be re-casted into
\begin{equation*}
\begin{split}
\mathop{{\rm max}}\limits_{(\cdot)} {\rm \; } \big \lbrace \vartheta_0(\mathscr{B}), \vartheta_1(\mathscr{B}), \vartheta_2(\mathscr{B}), \vartheta_3(\mathscr{B}), \vartheta_4^{\prime}, \vartheta_5^{\prime}, \vartheta_6^{\prime}\big \rbrace \;\;\;\; \; \; \; \; \; \; \; \\
\big \lbrace \vartheta_0(\mathscr{B}), \vartheta_1(\mathscr{B}), \vartheta_2(\mathscr{B}), \vartheta_3(\mathscr{B}), \vartheta_4^{\prime}, \vartheta_5^{\prime}, \vartheta_6^{\prime}\big \rbrace \le \; \; \; \; \; \; \; \; \\
\big \lbrace \vartheta_0(\mathscr{B}), \vartheta_1(\mathscr{B}), \vartheta_2(\mathscr{B}), \vartheta_3(\mathscr{B}), \cdots \;\;\;\;\;\;\;\;\;\;\;\;\;\;\;\;\;\;\;\;\;\;\;\\ \vartheta_4(\mathscr{B}), \vartheta_5(\mathscr{B}), \vartheta_6(\mathscr{B})\big \rbrace. \;\;\;
\end{split}
\end{equation*}

\textsc{Step 4.} \textcolor{black}{As some of the above-mentioned thresholds have still remained functions of $\mathscr{B}$}, even though the optimisation problem obtained above has now been relaxed further, it needs more relaxation.

As a case in point $-$ which is extendible to the other expressions $-$ , we know $\mathcal{P}{\big(\mathcal{Y}^{(o)}_q|\mathcal{X}_q} \big)$ needs $\mathcal{P}{\big(\mathcal{S}|\mathcal{Y}^{(o)}_q} \big)$ and $\mathcal{P}{\big(\mathcal{Y}^{(o)}_q} \big)$ as well, to be solved. We see that it is more probable to see the uncertainty in $\mathcal{P}{\big(\mathcal{S}|\mathcal{Y}^{(o)}_q} \big)$, i.e., $\mathcal{P}{\big(\mathcal{S}|\mathcal{Y}^{(o)}_q;\mathscr{B}} \big)$ compared with other cases $-$ chiefly arose from the feedback error or/and delays etc.

On the other hand, we know\footnote{\textcolor{black}{See e.g. \cite{2}.}} 
\begin{equation*}
\begin{split}
\mathcal{P}{\big(\mathcal{S}|\mathcal{Y}^{(o)}_q};\mathscr{B} \big)=\sum_{x}\mathcal{P}{\big(\mathcal{S}|\mathcal{X}_q} \big)\mathcal{P}{\big(\mathcal{X}|\mathcal{Y}^{(o)}_q};\mathscr{B} \big),
\end{split}
\end{equation*}
where 
\begin{equation*}
\begin{split}
\mathcal{P}{\big(\mathcal{X}|\mathcal{Y}^{(o)}_q};\mathscr{B} \big)=
\frac{\mathcal{P}{\big(\mathcal{X}_q} \big)}{\mathbb{Z}(\mathcal{Y}^{(o)}_q; \omega; \mathscr{B})}\;\;\;\;\;\;\;\;\;\;\;\;\;\;\;\;\;\;\;\;\;\;\;\;\;\;\;\;\;\\exp \Bigg \lbrace - \omega \bigg \lbrace \mathcal{P}{\big(\mathcal{S}|\mathcal{Y}^{(o)}_q} ;\mathscr{B}\big) || \mathcal{P}{\big(\mathcal{S}|\mathcal{X}_q} \big) \bigg \rbrace \Bigg \rbrace,
\end{split}
\end{equation*}
where $\mathbb{Z}(\cdot)$ is the normalization factor, also known as the partition function, and $\cdot||\cdot$ stands for the Kullback-Leibler divergence $-$ a.k.a with the relative entropy.

\textcolor{black}{Calling} the last mathematical expression, we see that
\begin{equation*}
\begin{split}
\mathop{{\rm max}}\limits_{(\cdot)} {\rm \; } \mathcal{P}{\big(\mathcal{S}|\mathcal{Y}^{(o)}_q};\mathscr{B} \big),
\end{split}
\end{equation*}
or
\begin{equation*}
\begin{split}
\mathop{{\rm max}}\limits_{(\cdot)} {\rm \; } \mathcal{P}{\big(\mathcal{X}|\mathcal{Y}^{(o)}_q};\mathscr{B} \big),
\end{split}
\end{equation*}
would be equivalent to
\begin{equation*}
\begin{split}
\mathop{{\rm max}}\limits_{(\cdot)} {\rm \; } \mathcal{P}{\big(\mathcal{S}|\mathcal{X}_q} \big),
\end{split}
\end{equation*}
in those cases that
\begin{equation*}
\begin{split}
\mathcal{P}{\big(\mathcal{S}|\mathcal{X}_q} \big) \le \mathcal{P}{\big(\mathcal{S}|\mathcal{Y}^{(o)}_q};\mathscr{B} \big),
\end{split}
\end{equation*}
holds $-$ whether or not due to $\mathscr{B}$.

In other words, we found a lower-bound for $\mathcal{P}{\big(\mathcal{S}|\mathcal{Y}^{(o)}_q};\mathscr{B} \big)$ which is not uncertainty included $-$ which is valid for the maximisation for at least some few cases.

\textsc{Step 5.} Meanwhile, we see that there inevitably exists\footnote{\textcolor{black}{Since as discussed e.g. in \cite{4}, $\mathcal{I} \big(\mathcal{S},\mathcal{X}_q\big) = \mathcal{H} \big(\mathcal{X}_q\big)-\mathcal{H} \big(\mathcal{S}|\mathcal{X}_q\big)$ where at the end of the curve, i.e., related to the information-bottleneck bound the second term goes to zero according to the deterministic feature. Now we see we do not have any term $\mathscr{B}$ here anymore.}} a pair $\big(\gamma^{\star}_2;\vartheta^{\star}_1 \big)$ where the condition 
\begin{equation*}
\begin{split}
\mathscr{Pr} \Big \lbrace\mathcal{I} \big(\mathcal{X}_q(\mathscr{B}),\mathcal{Y}^{(o)}_q(\mathscr{B})\big) \ge \gamma^{\star}_2 \Big \rbrace \ge \vartheta^{\star}_1,
\end{split}
\end{equation*}
makes $\mathcal{I} \big( \mathcal{Y}^{(o)}_q(\mathscr{B}),\mathcal{S}(\mathscr{B})\big)$ to be bounded towards the \textit{information-bottleneck-bound}\footnote{See e.g. \cite{4, 25} in order to understand what it is.} $\mathcal{I} \big(\mathcal{S},\mathcal{X}_q\big)$ $-$ something that shows that we have now found an additional relaxation over the main problem.

\textsc{Step 6.} Now, only 
\begin{equation*}
\begin{split}
\mathscr{Pr} \Big \lbrace \mathcal{I} \big(\mathcal{X}_q(\mathscr{B}),\big \lbrace\mathcal{Y}^{(tot)}_{q^{\prime}}(\mathscr{B})\big \rbrace\big) \le \gamma_3 \Big \rbrace \ge \vartheta_0,
\end{split}
\end{equation*}
remains un-relaxed. So, we continue as follows. We know 
\begin{equation*}
\begin{split}
\mathscr{Pr} \big \lbrace \mathcal{B} \cup \mathcal{C} \big \rbrace =\mathscr{Pr} \{ \mathcal{B} \} +\mathscr{Pr} \{ \mathcal{C} \}-\mathscr{Pr} \big \lbrace \mathcal{B} \cap \mathcal{C} \big \rbrace,
\end{split}
\end{equation*}
where $\mathscr{Pr} \big \lbrace \mathcal{B} \cap \mathcal{C} \big \rbrace$ is equivalent\footnote{\textcolor{black}{According to the definition of e.g. Kullback-Leibler divergence in relation to the latent variables as e.g. described for $\mathcal{P}{\big(\mathcal{X}|\mathcal{Y}^{(o)}_q};\mathscr{B} \big)$ in Step 4.}} to either $\mathscr{Pr} \big \lbrace \mathcal{B} |\mathcal{C} \big \rbrace$ or $\mathscr{Pr} \big \lbrace \mathcal{C} | \mathcal{B} \big \rbrace$, which should make any sense. Since 
\begin{equation*}
\begin{split}
\big \lbrace\mathcal{Y}^{(tot)}_q\big \rbrace=\Big \lbrace\mathcal{Y}^{(o)}_q;\big \lbrace\mathcal{Y}^{(v)}_{q}\big \rbrace \Big \rbrace=\Big \lbrace\mathcal{Y}^{(o)}_q \cup\big \lbrace\mathcal{Y}^{(v)}_{q}\big \rbrace \Big \rbrace,
\end{split}
\end{equation*}
holds, so 
\begin{equation*}
\begin{split}
\mathop{{\rm min}}\limits_{(\cdot)} {\rm \; }\mathcal{I} \Big(\mathcal{X}_q(\mathscr{B}),\big \lbrace\mathcal{Y}^{(tot)}_{q^{\prime}}(\mathscr{B})\big \rbrace\Big),
\end{split}
\end{equation*}
is equivalent to\footnote{\textcolor{black}{As described in details in Appendix \ref{sec:A}: privacy vs. quality-of-serice.}} 
\begin{equation*}
\begin{split}
\mathop{{\rm max}}\limits_{(\cdot)} {\rm \; } \mathcal{I} \Big(\mathcal{X}_q(\mathscr{B}),\big \lbrace\mathcal{Y}^{(v)}_{q^{\prime}}(\mathscr{B})\big \rbrace|\mathcal{Y}^{(o)}_{q^{\prime}}(\mathscr{B})\Big),
\end{split}
\end{equation*}
or
\begin{equation*}
\begin{split}
\mathop{{\rm min}}\limits_{(\cdot)} {\rm \; } \sum_{\mathcal{P} \big( \mathscr{y}^{(o)}_{q^{\prime}}(\mathscr{B}) \in \mathcal{Y}^{(o)}_{q^{\prime}}(\mathscr{B}) \big)} \bigg \lbrace \cdots\;\;\;\;\;\;\;\;\;\;\;\;\;\;\;\;\;\;\;\;  \\
 \mathcal{P} \Big(\mathcal{X}_q(\mathscr{B}),\big \lbrace\mathcal{Y}^{(v)}_{q^{\prime}}(\mathscr{B})\big \rbrace|\mathcal{Y}^{(o)}_{q^{\prime}}(\mathscr{B})\Big) \cdots \;\;\;\;\;\;\;\;\;\;\;\;\;\;\;\;\; \\ || \mathcal{P} \Big(\mathcal{X}_q(\mathscr{B})|\mathcal{Y}^{(o)}_{q^{\prime}}(\mathscr{B})\Big)\mathcal{P} \Big(\big \lbrace\mathcal{Y}^{(v)}_{q^{\prime}}(\mathscr{B})\big \rbrace|\mathcal{Y}^{(o)}_{q^{\prime}}(\mathscr{B})\Big)\bigg \rbrace,
\end{split}
\end{equation*}
or 
\begin{equation*}
\begin{split}
\mathop{{\rm max}}\limits_{(\cdot)} {\rm \; }\mathcal{P} \Big(\mathcal{X}_q(\mathscr{B})|\mathcal{Y}^{(o)}_{q^{\prime}}(\mathscr{B})\Big),
\end{split}
\end{equation*}
in those cases that 
\begin{equation*}
\begin{split}
\mathcal{P} \Big(\mathcal{X}_q(\mathscr{B}),\big \lbrace\mathcal{Y}^{(v)}_{q^{\prime}}(\mathscr{B})\big \rbrace|\mathcal{Y}^{(o)}_{q^{\prime}}(\mathscr{B})\Big) \ge\;\;\;\;\;\;\;\;\;\;\;\;\;\;\;\;\;\;\;\;  \\ \mathcal{P} \Big(\mathcal{X}_q(\mathscr{B})|\mathcal{Y}^{(o)}_{q^{\prime}}(\mathscr{B})\Big)\mathcal{P} \Big(\big \lbrace\mathcal{Y}^{(v)}_{q^{\prime}}(\mathscr{B})\big \rbrace|\mathcal{Y}^{(o)}_{q^{\prime}}(\mathscr{B})\Big),
\end{split}
\end{equation*}
holds $-$ whether or not due to $\mathscr{B}$. Finally, in order to calculate 
\begin{equation*}
\begin{split}
\mathcal{P} \Big(\big \lbrace\mathcal{Y}^{(v)}_{q^{\prime}}(\mathscr{B})\big \rbrace|\mathcal{Y}^{(o)}_{q^{\prime}}(\mathscr{B})\Big),
\end{split}
\end{equation*}
we need $\mathcal{P} \big(\mathcal{X}_{q}\big)$, $\mathcal{P} \Big(\mathcal{S}|\mathcal{Y}^{(o)}_{q^{\prime}}(\mathscr{B})\Big)$ as well as $\mathcal{P} \Big(\mathcal{S}|\mathcal{X}_{q}\Big)$, while all the three last terms are independent of $\mathscr{B}$ as e.g. we found a sp\textcolor{black}{e}cific case where $\mathcal{P} \Big(\mathcal{S}|\mathcal{X}_{q}\Big)$ could be relaxed, proven in some lines above $-$ in the Step 5.

The proof is now completed.$\; \; \; \blacksquare$

\section{Proof of Proposition 3}
\label{sec:C}
We use \textit{contradiction}. Initially speaking, we assume that there is no equilibrium. Conversely, we see that we have a Torus, or simply speaking, a two-nested-circle platform in a 2-D zone, with the radii $0<\rho_1<\rho_2$ and the region $\rho_1<\rho<\rho_2$ is valid since: (\textit{i}) on the one hand, we have to preserve the players as mush as we can in the coalition, (\textit{ii}) on the other hand, we have to constrain them for the approval of the total resources by removing them from the coalition. Thus, we can theoretically see that the region 
\begin{equation*}
\begin{split}
\rho^{\star}:=\{\rho| \rho_1<\rho<\rho_2\},
\end{split}
\end{equation*}
in the Torus defined above is an inevitable equilibrium and this is a contradiction.

Indeed, there exists a nested \textit{bi-level zero-sum} game in which one group of users lose the outer game whereas they undoubtedly win the inner game. So, although neither does the outer game have an equilibrium nor the inner game, the overall bi-level game falls in the region 
\begin{equation*}
\begin{split}
\rho^{\star}:=\{\rho| \rho_1<\rho<\rho_2\},
\end{split}
\end{equation*}
in the Torus defined above.

Additionally, one can interpret the obtained Nash equilibrium in the context of \textit{max $\mathbb{K}-$cut} game as follows. In a train for the given number of users, we have $\mathbb{K}$ Wagons, named \textit{colours} where the pay-off for the $i$th user would be 
\begin{equation*}
\begin{split}
\mathscr{P}_{\mathscr{off}}:=\mathop{{\rm max}}\limits_{(\cdot)} {\rm \; }\sum_{i, \mathscr{s}_i=\mathscr{s}_j, i \neq j}\mathscr{w}_{i \rightarrow j},
\end{split}
\end{equation*}
for the strategy pair $(\mathscr{s}_i, \mathscr{s}_i)$ and the weight $\mathscr{w}_{i \rightarrow j}$. So, with regard to $\varphi_1$ and $\varphi_2$ which respectively guarantee the \textit{inference v.s. privacy}, 
\begin{equation*}
\begin{split}
\mathscr{Pr} \big \lbrace \varphi_1 \le|\mathscr{w}_{i \rightarrow j}| \le \varphi_2 \big \rbrace,
\end{split}
\end{equation*}
strongly exposes the Torus expressed above which is a contradiction $-$ that is, a case which is in contrast to the \textit{initial assumption}.

This completes the proof.$\; \; \; \blacksquare$

\section{Proof of Proposition 4}
\label{sec:D}
We saw that we can construct $\mathscr{A}_i, i \in \{1, 2, 3\}$ in the context of the probabilities defined in Proposition 2. In addition to this, succinctly speaking, some of the aforementioned probabilities have an ascending trend whereas the remaining ones have a descending trend $-$ as implicitly symbolised in the proof of Proposition 3 \textcolor{black}{in relation to the Torus discussed above}\footnote{\textcolor{black}{For more details and about the rotations of the probabilities relating to each other see e.g. \cite{2} $-$ where an \textit{Alternating} optimisation method was proposed.}}. This controversial ascending and descending trends in two groups of probabilities make assure us about the joint stability-detectablity-controllablity-stabilisablty-observablity-and-detectablity.

The proof is now completed.$\; \; \; \blacksquare$
\section{Proof of Lemma 1}
\label{sec:E}
We are aware of the fact that we should let the following be satisfied \textcolor{black}{$-$ as rate-distortion theory entails}
\begin{equation*}
\begin{split}
\mathop{{\rm min}}\limits_{\mathcal{P}{\big(\mathcal{Y}|\mathcal{X}} \big)} {\rm \; } \mathcal{I} \big(\mathcal{Y},\mathcal{S}\big) \;\;\;\;\; \;\;\;\;\; \;\;\;\;\; \;\;\;\;\; \;\;\;\;\; \;\;\;\;\;\;\;\;\; \\ s.t. \;\; \mathcal{I} \big(   \mathcal{X},\mathcal{S} \big) -\mathcal{I} \big(   \mathcal{Y},\mathcal{S} \big) \longrightarrow \epsilon^{+}.
\end{split}
\end{equation*}
The term $\epsilon^{+}$ is also the non-zero \textit{distortion} threshold. We observe that $\mathscr{U}(k)$ is a function of: (\textit{i}) $\mathscr{U}_0(k)$; (\textit{ii}) $\mathcal{P}\big(\mathcal{Y}|\mathcal{X} \big)$ $-$ call it hereinafter $\mathcal{P}(k)$ $-$ ; as well as (\textit{iii}) $\hat{\mathcal{X}}(k)$. 

This completes the proof.$\; \; \; \blacksquare$

\section{Proof of Proposition 5}
\label{sec:F}
The proof is easy to follow by recalling \textit{Lemma 1}. Let us define the optimisation problem 
\begin{equation*}
\begin{split}
\mathop{{\rm \mathbb{M}ax}}\limits_{\mathscr{q}} | \mathscr{U}_{\mathscr{q}}\rangle \;\;\;\;\;\;\;\;\; \;\;\;\;\;\;\;\;\; \;\;\;\;\;\;\;\;\; \;\;\;\;\;\;\;\;\; \;\;\;\;\;\;\;\;\; \;\;\;\;\;\;\;\;\; \;\;\;\;\;\;\;\;\; \;\;\; \\
s.t. \;\; i \mathscr{h} \frac{\mathscr{d}}{\mathscr{d}k} | \mathscr{U}_{\mathscr{q}}\rangle=\mathscr{H}_{\mathscr{q}}| \mathscr{U}_{\mathscr{q}}\rangle+\alpha \sum_{\mathscr{q}^{\prime}} \cdots \;\;\;\;\;\;\;\;\; \;\;\;\;\;\;\;\;\; \;\;\;\;\;\\  \beta_{\mathscr{qq}^{\prime}} \big \lbrace | \mathscr{U}_{\mathscr{q}}\rangle- | \mathscr{U}_{\mathscr{q}^{\prime}}\rangle \langle \mathscr{U}_{\mathscr{q}} | \mathscr{U}_{\mathscr{q}^{\prime}}\rangle \big \rbrace, \mathscr{q} \in \{  \mathscr{U}_0, \mathscr{P}, \hat{\mathcal{X}}  \}, \mathscr{q}^{\prime} \neq \mathscr{q},
\end{split}
\end{equation*}
where the constraint is justified according to the Lohe model\footnote{See e.g. \cite{27} to understand what it is.}, while the Hamiltonian $\mathscr{H}$ is the overall averaged energy of all the oscillators. The parameter of $\mathscr{h}$ is its vectorised version, additionally, the parameter of $\alpha$ is also a constant.$\; \; \; \blacksquare$

\section{Proof of Proposition 6}
\label{sec:G}
The proof is provided in terms of the following two-step solution, prior to which Lemma 1 should either be initially recalled. 

One can now define a stochastic Stackelberg game\footnote{See e.g. \cite{28} to understand what it is.} in which the Leader $\mathscr{U}_0$ is a Major-Player who has a dominant effect on other minor players $\mathscr{U}_{\mathscr{k}^{\prime} }, k^{\prime} \neq0$ $-$ call the Followers. In this game, the Leader is non-stationary, so, one can apply a bi-level probabilistic optimisation method\footnote{See e.g. \cite{29} to understand what it is.}. 

\textsc{Step 1:} The Leader aims at obtaining $\mathcal{P} \big( \mathscr{U}_{\hat{\mathcal{X}}};\mathscr{U}_{\mathscr{U}_0} \big)$ by solving 
\begin{equation*}
\begin{split}
\mathop{{\rm \mathbb{M}ax}}\limits_{\mathcal{P} \in \mathcal{P} \big( \mathscr{U}_{\hat{\mathcal{X}}};\mathscr{U}_{\mathscr{U}_0} \big) } | \mathscr{U}_{\mathscr{q}}\rangle, \mathscr{q} \in \{  \mathscr{U}_0, \mathscr{P}, \hat{\mathcal{X}}  \}.
\end{split}
\end{equation*}

\textsc{Step 2:} The Follower-set aims to find $\mathscr{U}_{\hat{\mathcal{X}}}$ by solving 
\begin{equation*}
\begin{split}
\mathop{{\rm \mathbb{M}ax}}\limits_{\hat{\mathcal{X}}} \mathbb{E}_{\mathscr{U}_{\mathscr{U}_0}}  \big \{  \mathcal{P} \big( \mathscr{U}_{\hat{\mathcal{X}}};\mathscr{U}_{\mathscr{U}_0} \big) \big \}.
\end{split}
\end{equation*}

\textsc{\textbf{Remark 1:}} \textit{Of course we have ignored $\mathbb{E}_k \{ \cdot \}$ for the ease of notation. Moreover, Step 1 optimally solves the main problem in the fashion of a forward solution $-$ as the inner loop $-$ , while, Step 2 optimally finds the backward of the problem $-$ As the outer loop.}

This completes the proof.$\; \; \; \blacksquare$

\section{Proof of Proposition 7}
\label{sec:H}
The proof is provided in terms of the following multi-step solution.

\textsc{Step 1:} Let us the following parameters be satisfied: $\sigma d \mathscr{W}_k$ is the noise for our MFG relating to parameter uncertainties; the probability density function (PDF) of the game as $\mathcal{P}_{df}(k)$, and $\bar{\mathcal{P}}$ as the average.

\textsc{Step 2:} Call the value function 
\begin{equation*}
\begin{split}
\mathscr{J}:=\mathop{{\rm \mathbb{M}ax}}\limits_{\mathcal{P} \in \mathcal{P} \big( \mathscr{U}_{\hat{\mathcal{X}}};\mathscr{U}_{\mathscr{U}_0} \big) } | \mathscr{U}_{\mathscr{q}}\rangle, \mathscr{q} \in \{  \mathscr{U}_0, \mathscr{P}, \hat{\mathcal{X}}  \},
\end{split}
\end{equation*}
while there exists the following non-conservative control law
\begin{equation*}
\begin{split}
\partial_k \mathscr{U}_{\mathscr{q}}=+\mathcal{P} \in \mathcal{P} \big( \mathscr{U}_{\hat{\mathcal{X}}};\mathscr{U}_{\mathscr{U}_0} \big) +\mathscr{W}_k,
\end{split}
\end{equation*}
with regard to the random walk process $\mathscr{W}_k$.

\textsc{Step 3:} One can now theoretically write the Hamilton-Jacobi-Bellman (HJB) and the Fokker-Planck-Kolmogorov (FPK) equations as eq. (\ref{eq:1}).

\begin{figure*}
\begin{equation}\label{eq:1}
\begin{split}
\begin{cases}
HJB: \; \mathop{{\rm \mathbb{S}up}}\limits_{\mathcal{P} (\cdot) \ge 0} {\rm \; } \cdots\\\;\;\;\; \;\;\;\;\;\;\left \{ \mathscr{U}_{\mathscr{q}}(k)+ \partial _{\mathcal{\mathscr{U}_{\mathscr{q}}}}\mathscr{J}(k;\mathscr{U}_{\mathscr{q}}(k)) \int_{\mathscr{U}_{\mathscr{q}}}\mathcal{P}(k)\mu_k dk-\bar{\mathcal{P}}(k)  \right \}+\partial_k \mathscr{J}+{\sigma}^2 \partial ^2_{\mathscr{U}_{\mathscr{q}}\mathscr{U}_{\mathscr{q}}} \mathscr{J}=0, \\
FPK: \;\partial _k \mathcal{P}_{df}(k)=  \sigma^2 \partial ^2_{\mathscr{U}_{\mathscr{q}}\mathscr{U}_{\mathscr{q}}\;\;\;\;} \left \{ \mathcal{P}_{df} (k)\right\}+\partial _{\mathscr{U}_{\mathscr{q}}} \left \{ \mathcal{P}_{df} \int_{\mathscr{U}_{\mathscr{q}}}\mathcal{P}(k)\mu_k dk\right \}..
\end{cases}
\end{split}
\end{equation}
\end{figure*}


\textsc{Step 4:} One should write the following in order to further guarantee the hardware complexity\footnote{See e.g. \cite{3} for more discussions.}
\begin{equation*}
\begin{split}
\mathop{{\rm min}}\limits_{\Omega_1 , \Omega_2}     \left \{     \underbrace{\mathcal{P}_{df}(0)}_{\Theta(\Omega_1)} \left|\right|\underbrace{\mathcal{P}_{df}(k_0)}_{\Theta(\Omega_2)} \right\}_{K-L}, \forall k_0 \in (0,\infty),
\end{split}
\end{equation*}
where $\{ \cdot || \cdot \}_{K-L}$ stands for the information-theoretic metric of \textit{Kullback-Leibler divergence}.

The proof is now completed.$\; \; \; \blacksquare$

\section{Proof of Proposition 8}
\label{sec:I}
The proof is provided in terms of the following two-step solution.

\textsc{Step 1:} The term
\begin{equation*}
\begin{split}
\int_{\mathscr{U}_{\mathscr{q}}}\mathcal{P}(k)\mu_k dk,
\end{split}
\end{equation*}
can be relaxed as 
\begin{equation*}
\begin{split}
\mathcal{P}(t_0)\underbrace{\int_{\mathscr{U}_{\mathscr{q}}}\mu_k dk}_{\mu^{\prime}(k)},
\end{split}
\end{equation*}
for a constant $t_0$ according to the \textit{mean value theorem} for integrals $-$ since $\mathcal{P}(k)$ is continuous over the $k$- horizon, where $\mu^{\prime}(k)$ is the mean value.

\textsc{Step 2:} Finite-element-method\footnote{See e.g. \cite{26} to understand what it is.}, says that we can find a weight $\gamma_{\mathscr{v}}$ in the sense that we can relax the non-stationary MFG by minimising 
\begin{equation*}
\begin{split}
||  \mu^{\prime}(k+1)-\mu^{\prime}(k)||^2,
\end{split}
\end{equation*}
as much as possible, that is, the following mathematical expression over the time horizon $[0,\mathscr{K}]$ where $\mathscr{V}$ is the arbitary number of clusters\footnote{If, in totally different times, all the $\gamma_{\mathscr{v}}$ are selected as $0$ or $1$, the clusters are called deterministic, otherwise, if the $\gamma_{\mathscr{v}}$ are selected as in $[0 , 1]$, the clusters are called \textit{fuzzy}.}
\begin{equation*}
\begin{split}
 \mathop{{\rm \mathbb{M}in}}\limits_{\gamma_{\mathscr{v}}(k),\mu^{\prime}(k+1)} \sum_{\mathscr{v}=0}^{\mathscr{V}} \sum_{k=0}^{\mathscr{K}}\gamma_{\mathscr{v}}(k)||  \mu^{\prime}(k+1)-\mu^{\prime}(k)||^2,\\
 \sum_{\mathscr{v}=0}^{\mathscr{V}} \gamma_{\mathscr{v}}(k)=1,\gamma_{\mathscr{v}}(k) \ge 0.\;\;\;\;\;\;\;\;\;\;\;\;\;\;
\end{split}
\end{equation*}

It can be theoretically seen that the game is of a smooth nature. Therefore, regarding the smooth nature of our MFG, there undoubtedly exists a Nash-equilibrium. This approves that our system model is controllable-and-detectable.

The proof is now completed.$\; \; \; \blacksquare$

\section{Proof of  Proposition 9}
\label{sec:J}
The proof is provided in the context of the following multi-step solution.

\textsc{Step 1:} We define
\begin{equation*}
\begin{split}
\mathcal{I}\big(\mathcal{X},\mathcal{Y}|\mathcal{Z}\big):=\sum_{\mathscr{z} \in \mathcal{Z}}\mathcal{P}\big(\mathcal{Z}=\mathscr{z}\big)\;\;\;\;\;\;\;\;\;\;\;\;\;\;\;\;\;\;\;\;\;\;\;\;\;\;\;\;\\
 \bigg \lbrace\mathcal{P}\big(\mathcal{X},\mathcal{Y}|\mathcal{Z}\big) \; \Big | \Big  |  \;   \mathcal{P}\big(\mathcal{X}|\mathcal{Z}=\mathscr{z}\big) \mathcal{P}\big(\mathcal{Y}|\mathcal{Z}=\mathscr{z}\big)  \bigg \rbrace,
\end{split}
\end{equation*}
so, one can apply $\sum_{\mathscr{y} \in \mathcal{Y}}  $ over the both hand-sides as
\begin{equation*}
\begin{split}
\sum_{\mathscr{z} \in \mathcal{Z}}\mathcal{P}\big(\mathcal{Z}=\mathscr{z}\big)\;\;\;\;\;\;\;\;\;\;\;\;\;\;\;\;\;\;\;\;\;\;\;\;\;\;\;\;\;\;\;\;\;\;\;\;\;\;\;\;\;\;\;\;\;\;\;\;\;\;\;\;\;\;\; \\  \bigg   \lbrace \sum_{\mathscr{y} \in \mathcal{Y}}   \Big  \lbrace   \mathcal{P}\big(\mathcal{X},\mathcal{Y}|\mathcal{Z}\big) \;\Big | \Big  | \; \mathcal{P}\big(\mathcal{X}|\mathcal{Z}\big) \mathcal{P}\big(\mathcal{Y}|\mathcal{Z}\big)  \Big \rbrace \bigg \rbrace=\;\; \;\; \\
\sum_{\mathscr{z} \in \mathcal{Z}}\mathcal{P}\big(\mathcal{Z}=\mathscr{z}\big)\;\;\;\;\;\;\;\;\;\;\;\;\;\;\;\;\;\;\;\;\;\;\;\;\;\;\;\;\;\;\;\;\;\;\;\;\;\;\;\;\;\;\;\;\;\;\;\;\;\;\;\;\;\;\;  \\ \bigg   \lbrace     \sum_{\mathscr{y} \in \mathcal{Y}} \mathcal{P}\big(\mathcal{X},\mathcal{Y}|\mathcal{Z}\big) \; \Big | \Big  |  \; \sum_{\mathscr{y} \in \mathcal{Y}}  \mathcal{P}\big(\mathcal{X}|\mathcal{Z}\big) \mathcal{P}\big(\mathcal{Y}|\mathcal{Z}\big)  \bigg \rbrace=\\
\sum_{\mathscr{z} \in \mathcal{Z}}\mathcal{P}\big(\mathcal{Z}=\mathscr{z}\big)  \;\;\;\;\;\;\;\;\;\;\;\;\;\;\;\;\;\;\;\;\;\;\;\;\;\;\;\;\;\;\;\;\;\;\;\;\;\;\;\;\;\;\;\;\;\;\;\;\;\;\;\;\;\;\;\\ \bigg   \lbrace     \sum_{\mathscr{y} \in \mathcal{Y}} 1  \;\Big | \Big  |  \; \sum_{\mathscr{y} \in \mathcal{Y}}  \mathcal{P}\big(\mathcal{Y}|\mathcal{Z}\big)  \bigg \rbrace\;\;\;\;\;\;\;\;\;\;\;\;\;\;\;\;\;\;\;\;\;\;\;\;\;\;\;\;\;\;\;\;\\
\end{split}
\end{equation*}
in which $\mathcal{P}\big(\mathcal{Y}|\mathcal{Z}\big)$ entails
\begin{equation*}
\begin{split}
\frac{\mathcal{P} \big(\mathcal{Y} \big)}{\Omega \big (   \theta_0 ;   \mathcal{Z} \big)} exp \bigg   \lbrace   - \theta_0 \Big   \lbrace     \mathcal{P}\big(\mathcal{M}| \mathcal{Y}\big)  \;\Big | \Big  |  \;  \mathcal{P}\big(\mathcal{M}|\mathcal{Z}\big) \Big   \rbrace \bigg \rbrace,
\end{split}
\end{equation*}
w.r.t. the arbitary $\mathcal{M}$\footnote{\textcolor{black}{As a latent one: This is an information theoretic trick in this context about which the interested reader should refer to e.g. \cite{22}.}}, where 
\begin{equation*}
\begin{split}
\; \mathcal{P}\big(\mathcal{Y}\big)=\sum_{\mathscr{x} \in \mathcal{X}} \mathcal{P}\big(\mathcal{X}\big)\mathcal{P}\big(\mathcal{Y}|\mathcal{X}\big),
\end{split}
\end{equation*}
in which $\mathcal{P}\big(\mathcal{Y}|\mathcal{X}\big)$ entails
\begin{equation*}
\begin{split}
\frac{\mathcal{P} \big(\mathcal{Y} \big)}{\Omega \big (   \theta_3 ;   \mathcal{X} \big)} exp \bigg   \lbrace   - \theta_3 \Big   \lbrace     \mathcal{P}\big(\mathcal{M}| \mathcal{Y}\big)  \;\Big | \Big  |  \;  \mathcal{P}\big(\mathcal{M}|\mathcal{X}\big) \Big   \rbrace \bigg \rbrace.
\end{split}
\end{equation*}

\textsc{Step 2:} The \textit{divergence theorem} indicates that 
\begin{equation*}
\begin{split}
\int_{\mathscr{V}}\big(\nabla \cdot \mathscr{F}\big)d\mathscr{V}=\oint_{\mathscr{S}}\big(\mathscr{F} \cdot \hat{n}\big)d\mathscr{S}.
\end{split}
\end{equation*}
This means if the volume $\mathscr{V}$ is partitioned into separate parts, the sum of the flux out of each component volume is physically equal to the flux out of the original volume. This theoretically means that for the overall divergence, we should go over each component’s divergence, the superposition of which gets us the result. Now, one can see that ${\mathcal{P} \big(\mathcal{Y} |\mathcal{Z}\big)} $ and ${\mathcal{P} \big(\mathcal{X} |\mathcal{Z}\big)}$ are crucial here. We examine if there exist any case in which ${\mathcal{P} \big(\mathcal{Y} |\mathcal{Z}\big)} $ and ${\mathcal{P} \big(\mathcal{X} |\mathcal{Z}\big)}$ behave in contrast to each other. Thus, let us examine if $\Big   \lbrace     \mathcal{P}\big(\mathcal{Y}|\mathcal{Z}\big)  \;\Big | \Big  |  \;  \mathcal{P}\big(\mathcal{X}|\mathcal{Z}\big) \Big   \rbrace$ is computable, i.e., if they are differentiable in relation to each other, for which we see the following differentiability   
\begin{equation*}
\begin{split}
log \frac{\mathcal{P} \big(\mathcal{Y} |\mathcal{Z}\big)}{\mathcal{P} \big(\mathcal{X} |\mathcal{Z}\big)}     \longrightarrow \;\;\;\;\;\;\;\;\;\;\;\;\;\;\;\;\;\;\;\;\;\;\;\;\; \;\;\;\; \;\;\;\; \;\;\;\; \;\;\;\; \;\;\;\; \;\;\;\;  \\    \frac{\bigg   \lbrace   - \theta_1 \Big   \lbrace     \mathcal{P}\big(\mathcal{U}| \mathcal{Y}\big)  \;\Big | \Big  |  \;  \mathcal{P}\big(\mathcal{U}|\mathcal{Z}\big) \Big   \rbrace \bigg \rbrace}{\bigg   \lbrace   - \theta_2 \Big   \lbrace     \mathcal{P}\big(\mathcal{M}| \mathcal{X}\big)  \;\Big | \Big  |  \;  \mathcal{P}\big(\mathcal{M}|\mathcal{Z}\big) \Big   \rbrace \bigg \rbrace}.
\end{split}
\end{equation*}
Now, according to e.g. \cite{30}, one can see that we can partition the divergence examination for two multiplied functions in an infinite number of intervals for which we see that there exists some cases where either both of the aforementioned functions are non-increasing or both are non-decreasing. This means that the divergence examination claimed above is followable for some cases in an acceptable fashion.

The proof is now completed.$\; \; \; \blacksquare$

\section{Proof of  Proposition 10}
\label{sec:K}

We define the following w.r.t. the inaccessible terms\footnote{\textcolor{black}{The terms $(\cdot)^{(i)}$ and $(\cdot)^{(-i)}$ stand respectively for the accessible and inaccessible terms $-$ relating to $\mathscr{B}$.}} $\mathcal{Z}^{(-i)}$
\begin{equation*}
\begin{split}
\mathop{{\rm max}}\limits_{(\cdot)} {\rm \; }\mathcal{I}\big(\mathcal{X},\mathcal{Y}|\mathcal{Z}^{(i)}\mathcal{Z}^{(-i)}\big):=\;\;\;\;\;\;\;\;\;\;\;\;\;\;\;\;\;\;\;\;\;\;\;\;\;\;\;\;\;\;\;\;\;\;\;\;\;\;\;\;\;\;\;\;\;\;\;\\
\sum_{\mathscr{z} \in \mathcal{Z}}\mathcal{P}\big(\mathcal{Z}^{(i)}=\mathscr{z}^{(i)} , \mathcal{Z}^{(-i)}=\mathscr{z}^{(-i)}  \big) \;\;\;\;\;\;\;\;\;\;\;\;\;\;\;\;\;\;\\ 
\bigg \lbrace\mathcal{P}\big(\mathcal{X},\mathcal{Y}|\mathcal{Z}^{(i)}\mathcal{Z}^{(-i)}\big) \; \Big | \Big  |  \;   \mathcal{P}\big(\mathcal{X}|\mathcal{Z}^{(i)}=\mathscr{z}^{(i)},\mathcal{Z}^{(-i)}=\mathscr{z}^{(-i)}\big) \cdots \\
\mathcal{P}\big(\mathcal{Y}|\mathcal{Z}^{(i)}=\mathscr{z}^{(i)},\mathcal{Z}^{(-i)}=\mathscr{z}^{(-i)}\big)  \bigg \rbrace,\;\;\;\;\;\;\;\;\;\;\;\;\;\;\;\;\;\;\\
s.t.\;\gamma_1 \le\mathbb{E}_{\mathscr{z}^{(-i)}}    \Big \lbrace  \mathcal{H}\big(\mathcal{M}|\mathcal{Z}^{(-i)}\big)  \Big \rbrace \le \gamma_2,\;\;\;\;\;\;\;\;\;\;\;\;\;\;\;
\end{split}
\end{equation*}
where the two arbitrary thresholds $\gamma_1$ and $\gamma_2$ can realise the \textit{Kinship principle} \cite{31} and \textit{chance constraints}.

Indeed, the equivocation based constraint proves that we can acceptably follow the amount of the differentiability of the surfaces over a Riemann-manifold \cite{32, 33}.

The proof is now completed.$\; \; \; \blacksquare$

\begin{IEEEbiography}[{\includegraphics[width=1in,height=1.25in,clip,keepaspectratio]{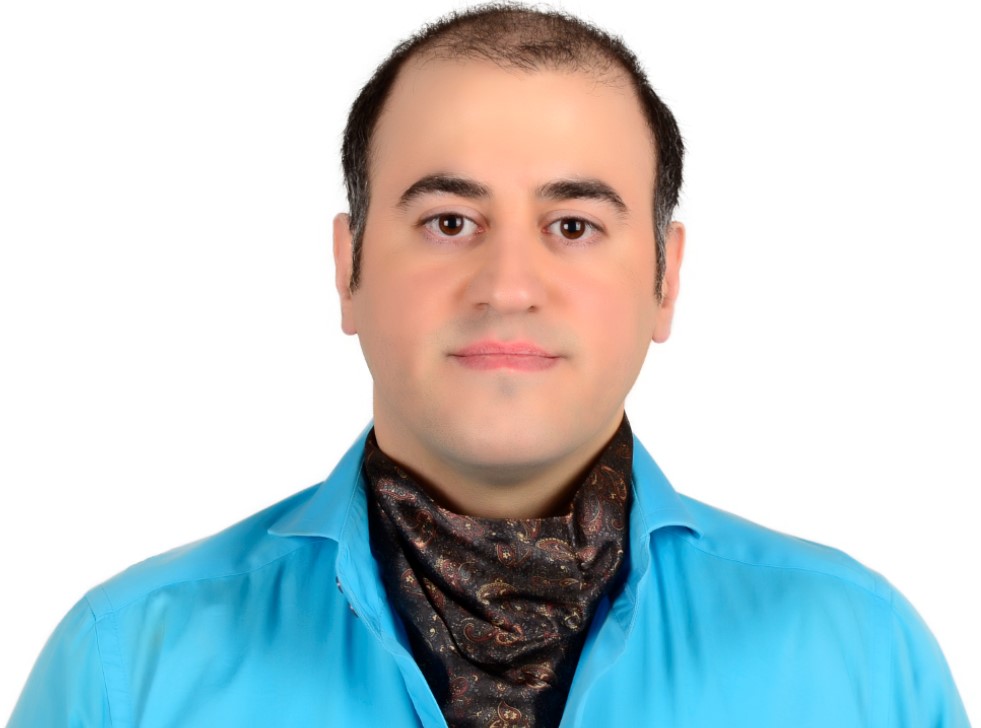}}]{Makan Zamanipour}
MAKAN ZAMANIPOUR (Researcher-ID: P-6298-2019; ORCID: 0000-0003-1606-9347; Scopus-ID: 56719734800) IEEE Member since 2015, born in Iran on 1983. His main research-field is Wireless communication theory, Information theory, Game theory and Optimisation. He has published a lot of papers in ISI-indexed journals as wll as reviewing for high-prestige ISI-indexed journals in IEEEs, Elsevier etc. His Google-Scholar profile and Publons are available online.
\end{IEEEbiography}
\markboth{Journal of \LaTeX\ Class Files,~Vol.~xx, No.~x, Xxxx~ 2021}%
{Shell \MakeLowercase{\textit{et al.}}: Bare Demo of IEEEtran.cls for Computer Society Journals}
\end{document}